\documentclass[preprint,aps,pre,nofootinbib,showpacs,showkeys]{revtex4}   
\usepackage{amsmath}
\usepackage{graphicx} 
\usepackage{amssymb}

\begin{document}

\title{Lyapunov modes in three-dimensional Lennard-Jones fluids}

\author{M. Romero-Bastida}

\affiliation{Facultad de Ciencias, Universidad Aut\'onoma del
 Estado de Morelos, Avenida Universidad 1001, Chamilpa, Cuernavaca, Morelos 62209, M\'exico}
\altaffiliation[Also at: ] {Departamento de F\'\i sica, Universidad Aut\'{o}noma Metropolitana
Iztapalapa, Apartado Postal 55--534, Distrito Federal 09340, M\'{e}xico}

\email{rbm@xanum.uam.mx}

\author{E. Braun}

\affiliation{Departamento de F\'\i sica, Universidad Aut\'{o}noma Metropolitana Iztapalapa,
Apartado Postal 55--534, Distrito Federal 09340, M\'{e}xico}

\date{\today{}}

\begin{abstract}
Recent studies on the phase-space dynamics of a one-dimensional Lennard-Jones
fluid reveal the existence of regular collective perturbations associated
with the smallest positive Lyapunov exponents of the system, called
hydrodynamic Lyapunov modes, which previously could only be identified
in hard-core fluids. In this work we present a systematic study of
the Lyapunov exponents and Lyapunov vectors, i.e. perturbations along
each direction of phase space, of a three-dimensional Lennard-Jones
fluid. By performing the Fourier transform of the spatial density
of the coordinate part of the Lyapunov vector components and then
time-averaging this result we find convincing signatures of longitudinal
modes, with inconclusive evidence of transverse modes for all studied
densities. Furthermore, the longitudinal modes can be more clearly identified
for the higher density values. Thus, according to our results, the
mixing of modes induced both by the dynamics and the dimensionality
induce a hitherto unknown type of order in the tangent space of the
model herein studied at high density values.
\end{abstract}

\pacs{05.45.Jn, 05.45.Pq, 02.70.Ns}

\keywords{Nonlinear dynamics, Lyapunov spectrum, Lyapunov modes, Fluid dynamics}

\maketitle

\section{Introduction\label{sec:Introduction}}

In recent years it has been possible to study in detail the underlying
chaotic dynamics of many-particle systems, thus finding interesting
connections between their microscopic dynamics and the observed macroscopic
behavior. For example, in the case of static properties, the largest
Lyapunov exponent (LLE), which measures the dynamical instability of
phase-space trajectories to infinitesimal perturbations in the initial
conditions, is related to the Kosterlitz-Thoules transition temperature in a
system of coupled rotators with nearest-neighbor interactions \cite{KT},
as well as to first \cite{grav1storder} and second-order phase transitions \cite{phtr2ndorder}
in particle systems with long-range interactions.
For dynamical properties, the sudden change in the gradient of the
LLE against energy, which corresponds to the transition from
weak to strong chaos, can be detected in the macroscopic behavior
of a Brownian particle coupled to the system \cite{Brownian}. It
has also been related to transport coefficients of fluid systems with
continuous potentials in non-equilibrium situations \cite{DC1}.
In equilibrium, a relation has also been proposed \cite{DC2}, although for this case
arguments both against \cite{CommBarnett} and in favor \cite{ReplyBarnett} have been advanced.
Nevertheless, these last results entail interesting
possibilities, since they suggest that dynamical instability is at
the origin of macroscopic transport phenomena. Two different methodologies
have related the Lyapunov spectrum (LS), defined as the sorted set
of Lyapunov exponents (LEs) which give the exponential rate of expansion or contraction
of nearby trajectories along each independent component of the phase
space, to the transport coefficients of simple fluids \cite{Method1,Method2}.
However, they involve setting up non equilibrium simulations or locating
special phase space trajectories. Furthermore, neither of these approaches
considers the perturbations in
phase space underlying the LEs, which have attracted a lot of attention
in recent years. Therefore it would be desirable first to attain a
more thorough understanding of the phase-space perturbations associated
to the LEs of the most general type of dynamical models employed to
study fluid systems before attempting the construction of a general theory that could
relate, with enough confidence, the dynamical instability with the
macroscopic behavior as described by the transport coefficients.

For some particular many-particle systems, e.g. hard-sphere fluids,
the theory of LEs is highly developed \cite{JRD}. Nevertheless, their
most interesting features have been discovered by means of molecular
dynamics simulations, which revealed that the slowly-growing and
decaying perturbations associated with the non-vanishing LEs
closest to zero are related to, and in some cases may even be
represented as, almost exact periodic vector fields coherently spread out over 
the physical space with well defined wave vectors,
 together with an almost precise stepwise structure of the considered LEs \cite{Milanovic}.
From their similarity to the classical modes of fluctuating hydrodynamics,
these perturbations have been named {\sl Lyapunov modes} (LMs). Since their
discovery \cite{OriginalLM}, much work has been done to understand
their origin and possible relevance. Some analytical 
approaches that have been advanced to understand these small LEs include
random-matrix dynamics \cite{EckmannGat,TaMoRM}, periodic orbit models
\cite{PeriodicOrbit}, and kinetic theory \cite{Mareschal,deWijn}. All
these approaches have met with only a very limited success, since
none of them has been able to satisfactorily describe the mode
dynamics in all the known situations accessible to simulations. 
A more serious drawback is that neither has been capable to extract 
the transport coefficients from the LMs, or to relate them,
in a simple way, to hydrodynamic fluctuations.

Until recently the existence of these LMs could only be verified in one, two and three-dimensional hard-sphere fluids \cite{OriginalLM}; their existence in the case of atomic fluids with soft interactions (both attractive and repulsive) remained controversial \cite{Hoover}, since certain features, such as the step structure in the LS, dissappear in soft-potential systems. Hence, to extend the concept of LMs beyond the realm of hard-core systems, a more generic definition has to be adopted, which involves both the appearance of a sharp peak of low wave number in the spatial Fourier spectrum of LVs corresponding to the LEs closest to zero and of a minimum in the average spectral entropy, defined in Sec. \ref{sec:TIIQ} of the present work, in that same LE region. Employing this more generic definition convincing evidence of the existence of the so defined LMs was obtained for some soft-potential systems, namely in fluids with a one-dimensional (1D) Lennard-Jones (LJ) \cite{Radons}, where the aforementioned definition  of LMs and the spectral analysis techniques needed to detect them were developed, and two-dimensional (2D) Weeks-Chandler-Anderson (WCA) interaction potentials \cite{WCA}. More recently, the existence of LMs has been also verified in lattices of coupled Hamiltonian and dissipative maps \cite{RadonsPRL,RadonsHLMCML,RadonsDynBehavHLMCML}, as well as in Fermi-Pasta-Ulam anharmonic oscillator lattices \cite{RadonsFPU}. Thus all information so far available seems to indicate that the Hamiltonian structure, conservation laws, and translational invariance are not necessary conditions for the existence of LMs. Nevertheless, since there is no theoretical scheme that can predict the existence of LMs, the discovery and characterization of LMs has to be done in a case-by-case basis. For example, the introduction of damping in a system of coupled Hamiltonian maps does not destroy the LMs, whereas the addition of that same damping to a system of coupled circle maps wipes off the LMs \cite{RadonsHLMCML}. Therefore, although LMs are present in the aforementioned systems, their existence is in no way guaranteed in other situations not studied so far, such as in three-dimensional Hamiltonian fluid systems with both attractive and repulsive interactions.
  
In this paper we perform a study of the dynamical instabilities of
a three-dimensional (3D) atomic fluid interacting with the full LJ potential
under the simulation conditions most frequently encountered in molecular
dynamics studies \cite{AT}. We assess the validity of various dynamical
indicators previously proposed in the literature for our particular
system. The main result of our work is that, although the spectral
analysis methods proposed in Ref.\cite{Radons} can be readily applied
to our system, the dimensionality greatly enhances the mixing among
perturbations that was already present in the case of the 1D LJ gas,
thus rendering both the detection and characterization of the collective
perturbations even more problematic.

This paper is organized as follows: In Sec. \ref{sec:The-Model} we
present a survey of the model, as well as the computational details
needed to obtain our results in the phase space of our system as well
as a short account of the relevant theory necessary to study the perturbations
of the phase space. Sec. \ref{sec:Lyapunov-Spectrum} describes the
results for the complete LS, its dependence on the particle density,
and provides evidence of the mixing among perturbations. In Sec.
 \ref{sec:spatial-structure-of-perturbations}
we show the results of the spectral analysis methods applied to the
proposed dynamical indicators. Section \ref{sec:discussion} is devoted
to discuss and analyze the results reported in the previous sections.
 In the Appendix we report some preliminary results for the temporal
 correlations of the phase space perturbations.
We present our conclusions in Sec. \ref{sec:conclusions}.

\section{The Model and simulation details\label{sec:The-Model}}

\subsection{Phase-space dynamics}

The LJ potential $U_{_{LJ}}(r)$ between particles $i$ and $j$ is
given by
\begin{equation}
U_{_{LJ}}(r)=4\epsilon\left[\left(\frac{\sigma}{r}\right)^{12}-\left(\frac{\sigma}{r}\right)^{6}\right],\label{eq:potLJ}\end{equation}
where $r\equiv r_{ij}$ is the distance between particles $i$ and
$j$; here $\sigma$ and $\epsilon$ are the LJ atom diameter and
the strength of the interparticle interaction, respectively. The actual
interaction potential employed in our simulations is the spherically
truncated and shifted (STS) potential, which can be written as
\begin{equation}
U(r)=\left\{ \begin{array}{cc}
U_{_{LJ}}(r)-U_{_{LJ}}(r_{c}) & r\le r_{c}\\
0 & r>r_{c},\end{array}\right.\label{eq:STS}
\end{equation}
 where $r_{c}=2.5\sigma$ is the cut-off radius at which the potential
is truncated in order to save computer time. It is important to notice
that the force derived from the above potential, as well as its derivative,
are not continuous at the truncation point $r_{c}$. Therefore the following
potential, first proposed in Ref. \cite{SFord}, can be employed
\begin{equation}
U(r)=\left\{ \begin{array}{cc}
4\epsilon\left[\left(\frac{\sigma}{r}\right)^{12}-\left(\frac{\sigma}{r}\right)^{6}\right] + c_2\left(\frac{r}{r_c}\right)^2 + U_{c} & r\le r_{c}\\
0 & r>r_{c},\end{array}\right.\label{eq:SF}
\end{equation}
 with $c_2=4\epsilon[6(\sigma/r_c)^{12}-3(\sigma/r_c)^{6}]$ and 
 $U_c=4\epsilon[-7(\sigma/r_c)^{12}+4(\sigma/r_c)^{6}]$. This potential
 has a continuous derivative in the truncation point $r_c$.
Nevertheless, for the 1D LJ model it has been shown, by adopting the above
expression for the potential, that the qualitative
behavior of the LMs is not greatly affected by the discontinuity of
the STS potential at $r_c$ \cite{Radons}. In the next section it will be 
 shown explicitly that this same phenomenology holds for our 3D LJ fluid.
The complete Hamiltonian of our system is then\begin{equation}
H=\sum_{i=1}^{N}\frac{p_{i}^{2}}{2m_{i}}+\sum_{i<j}U(r_{ij}),\label{eq:HamLJ}\end{equation}
with $\left\{ \mathbf{p}_{i}\right\} $ being the momenta of the atoms
and $\left\{ m_{i}\right\} $ their corresponding masses. In all our
simulations we took the masses of the atoms equal to unity, i.e. $m_{i}=m=1\:\forall\: i$,
with $\sigma=1$, and $\epsilon=1$ as well. The Hamiltonian is then
written in terms of the reduced variables $r^{*}=r/\sigma,$ $p_{i}^{*}=p_{i}/\left(m\epsilon\right)^{1/2}$,
and $m_{i}^{*}=m_{i}/m$.

The initial configuration for all simulations was set up from a fcc lattice on a square cell of sides $L^{*}=L_{x}^{*}=L_{y}^{*}=L_{z}^{*}$. The initial momenta were drawn from a Maxwell-Boltzmann distribution. Then the $6N$ equations of motion were numerically integrated by means of the Verlet leap-frog algorithm with periodic boundary conditions and the minimum image convention applied in all directions. A time-step of $\Delta t^{*}=\Delta t\left(\epsilon/m\right)^{1/2}/\sigma=0.001$ was used in all simulations, with an equilibration period of $100\,\,000$ steps at constant temperature obtained by a uniform rescaling of the velocities at each time-step. After the equilibration period the system was allowed to evolve at constant total energy for a period of $10^{6}$ time-steps. The relative energy drift for this number of time-steps is $10^{-4}$--$10^{-5}$, which is an acceptable compromise between accuracy and speed, since there is no systematic drift and thus the energy fluctuations are stable for the chosen time-step. All the reported results were obtained for systems of $N=108$ at a supercritical reduced temperature of $T^{*}=k_{_{B}}T/\epsilon=1.5$, where $k_{_{B}}$ is the Boltzmann constant. The reduced densities $\rho^{*}=\rho\,\sigma^{3}$ were taken within the range $\rho^{*}\in\left[0.01,0.5\right]$. The critical temperature and density for the LJ fluid with STS potential are $T_c^{*}=1.085$ and $\rho_c^{*}=0.317$ \cite{BSmit}; thus the employed values for these variables ensure that the system state is far away from the two-phase region in the $\rho^{*}$ vs $T^{*}$ phase diagram, being a homogeneous fluid. Finally, since in the rest of the work reduced variables will be exclusively employed, from now on we will drop the asterisk from all symbols without risk of confusion.

\subsection{Tangent-space dynamics}

The phase space trajectory is represented by the variable $\mathbf{\Gamma}(t)\equiv\left(\mathbf{\Gamma}_{1}\left(t\right),\mathbf{\Gamma}_{2}\left(t\right),\ldots,\mathbf{\Gamma}_{N}\left(t\right)\right)$,
where $\mathbf{\Gamma}_{i}\left(t\right)\equiv\left(\mathbf{r}_{i}\left(t\right),\mathbf{p}_{i}\left(t\right)\right)$.
To study the local dynamical stability of our system we introduce
the \emph{Lyapunov vector} (LV) as $\delta\mathbf{\Gamma}^{\left(\alpha\right)}(t)\equiv\left(\delta\mathbf{\Gamma}_{1}^{\left(\alpha\right)}\left(t\right),\delta\mathbf{\Gamma}_{2}^{\left(\alpha\right)}\left(t\right),\ldots,\delta\mathbf{\Gamma}_{N}^{\left(\alpha\right)}\left(t\right)\right)$, where $\alpha=1,\ldots,6N$ and
$\delta\mathbf{\Gamma}_{i}^{\left(\alpha\right)}\left(t\right)\equiv\left(\delta\mathbf{r}_{i}^{\alpha}\left(t\right),\delta\mathbf{p}_{i}^{\alpha}\left(t\right)\right)$
representing the \emph{i}th particle contribution to the infinitesimal
perturbations of the trajectory $\mathbf{\Gamma}\left(t\right)$ along
all possible directions (position and momentum axes) of the phase
space, thus defining the so called \emph{tangent space}. 

According to Oseledec's multiplicative ergodic theorem \cite{Oseledec} the remote past limit symmetric operator ${\Phi}_{\mathrm b}(t)=\lim_{t_0\rightarrow-\infty}[{\cal M}(t,t_0)\cdot{\cal M}^{\mathrm T}(t,t_0)]^{1/[2(t-t_0)]}$ exists for almost every initial condition $\mathbf{\Gamma}\left(t_0\right)$, where ${\cal M}(t,t_0)$ is the fundamental matrix governing the time evolution of the perturbations $\delta\mathbf{\Gamma}\left(t\right)$ in tangent space as $\delta\mathbf{\Gamma}(t)={\cal M}(t,t_0)\cdot\delta\mathbf{\Gamma}(t_0)$ \cite{Method2}. The set of instantaneous LEs is defined as $\lambda^{(\alpha)}(t)=\ln\Lambda^{(\alpha)}$, where $\Lambda^{(\alpha)}$ are the eigenvalues of $\Phi_{\mathrm b}(t)$. The herein employed standard procedure to compute the LEs consists in periodically reorthonormalizing a set of offset vectors that are time evolved by means of the matrix ${\cal M}(t,t_0)$ \cite{Benettin,Shimada}. The time-averaged values of the logarithms of the renormalization factors, i.e. $\langle\ln\Lambda^{(\alpha)}\rangle_t$ are the LEs $\left\{ \lambda^{\left(\alpha\right)}\right\}$ and the set of offset vectors right after the reorthonormalization are the eigenvectors of $\Phi_{\mathrm b}(t)$, which are called \emph{backward} LVs. The equivalence of the LEs computed by means of these two methods can be proved rigorously \cite{Johnson}, but the relation between the Oseledec eigenvectors and the LVs obtained via the standard method is more subtle. It is known that the backward LVs converge at an exponential rate to the Oseledec eigenvectors for the inverse-time dynamics of the original system \cite{Orszag,Ershov}. The latter are obtained as the eigenvectors of the far future limit operator ${\Phi}_{\mathrm f}(t)=\lim_{t_0\rightarrow\infty}[{\cal M}^{\mathrm T}(t_0,t)\cdot{\cal M}(t_0,t)]^{1/[2(t_0-t)]}$ and are called \emph{forward} LVs. It is to be noted that recently it has been possible to obtain from the intersection of the embedded subspaces spanned by the eigenvectors of ${\Phi}_{\mathrm b}(t)$ (backward LVs) and ${\Phi}_{\mathrm f}(t)$ (forward LVs) the so called \emph{characteristic} LVs, which are independent of the norm and do not form an orthogonal basis \cite{Diego}. A somewhat similar algorithm has also been proposed in Ref. \cite{Ginelli}. The aforementioned schemes are based on ideas already discussed long ago \cite{ER}, but have only been recently proposed because their implementation is by no means a simple task from a computational point of view, which is why they only have been tested in simple 1D systems. Thus it is not at all clear that their implementation could be feasible in the near future for the case of our system. Furthermore, since the purpose of the present work is to investigate the possible existence of LMs in a 3D LJ fluid, it is important to recall that the discovery of LMs in all studied systems so far has been made employing the backward LVs. So it seems that the use of the CLVs is not essential for the detection of LMs. Finally, meaningful results are still obtained by means of the backward LVs obtained from the standard procedure, such as in the recent characterization of LMs in a diatomic system \cite{Diatomic} and the discovery of LMs in the XY rotator model \cite{XYmodel}; in both studies the backward LVs were employed. Therefore, from now on we will refer to the numerically computed vectors (backward LVs) as the LVs without confusion.
 
In the Hamiltonian case (which we treat here), and also in some special
homogeneous non-equilibrium situations \cite{cpr1}, the LEs and the
corresponding LVs have a symmetry property which makes unnecessary
to calculate the whole spectrum. In these cases the LS thus computed
is symmetrical around zero, which means that each LE has a partner
that is exactly its negative. This is the so called \emph{conjugate-pairing
rule} \cite{cpr2}. Therefore, only $3N$ linearized equations for
the LVs were simultaneously integrated, along with the $6N$ nonlinear
equations for the reference trajectory $\mathbf{\Gamma}(t)$. The initial
 LVs $\delta\mathbf{\Gamma}^{\left(\alpha\right)}\left(t_0\right)$
in all our simulations consisted in a set of $3N$ orthogonal vectors
with randomly selected components.

\section{Lyapunov Spectrum \label{sec:Lyapunov-Spectrum}}

\begin{figure}
\includegraphics[%
  scale=0.3]{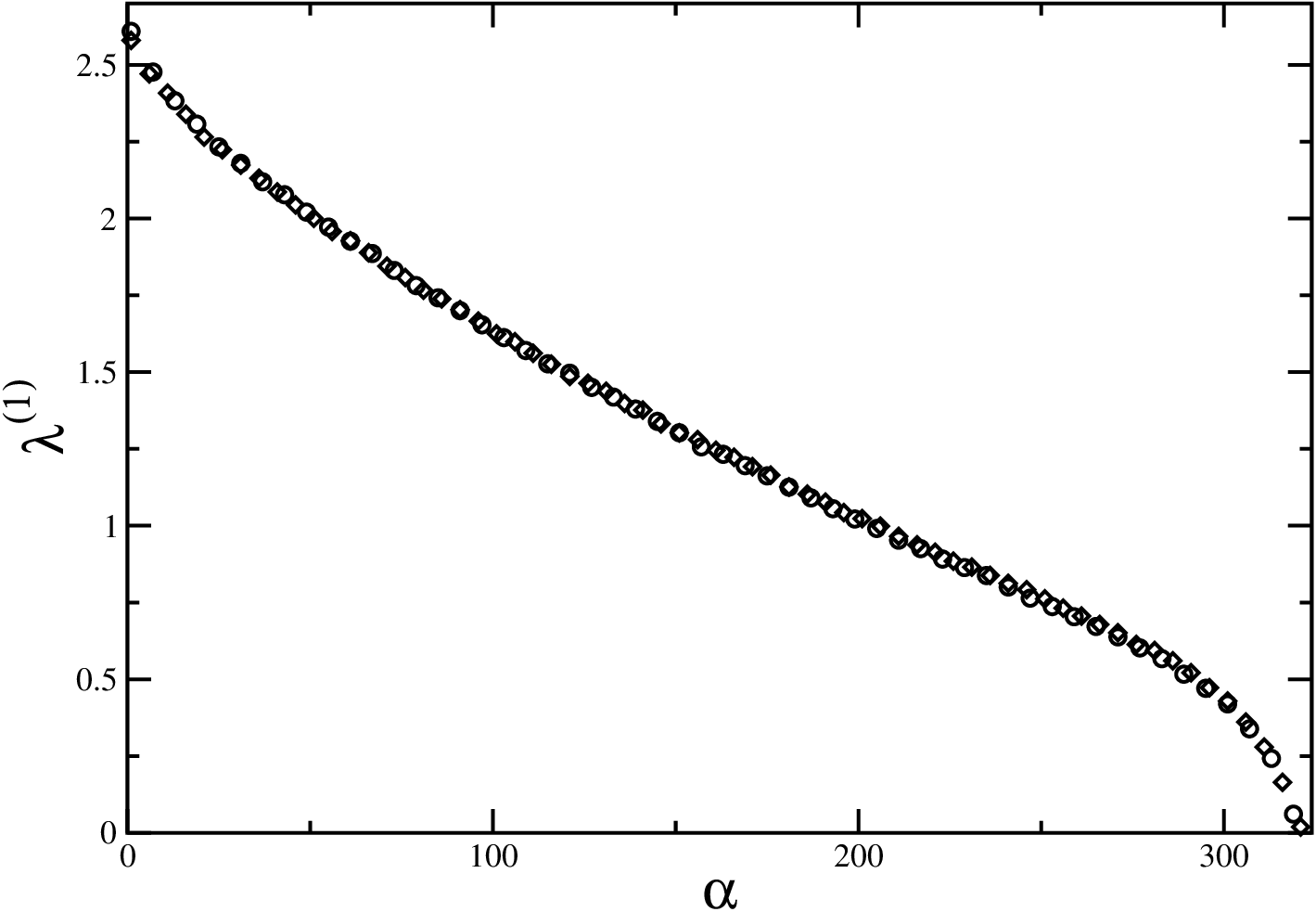}
\caption{\label{cap:LS-STS-SF} LS $\left\{ \lambda^{\left(\alpha\right)}\right \}$ as a function of the Lyapunov index $\alpha$ computed for a system with the STS potential (circles) and the potential defined by Eq. (\ref{eq:SF}) (diamonds). The thermodynamic state in both cases is defined by $\rho=0.1$ and $T=1.5$, with $N=108$.}
\end{figure}

As mentioned in the previous section, both the force derived from
 the STS potential, as well as its derivative, are not continuous at $r_c$.
 In order to assess the relevance of this particular feature in the
 computation of the properties in the tangent space we perform additional
 simulations employing the potential given by Eq. (\ref{eq:SF}).
 In Fig. \ref{cap:LS-STS-SF} we
 present the LS computed from this last potential, as well as that computed by
 means of the STS potential, for $T=1.5$, $\rho=0.1$, and $N=108$. As 
 can be readily appreciated there are no significant differences,
 a result which supports the use of the simpler STS potential.
 The most relevant feature of this figure is that,
 as in the case of the 1D LJ gas, in the smallest positive
LE region there is no evidence of the stepwise structure that signals
the appearance of the LMs in the case of hard-core systems.
Thus it is plausible that the same mechanism that accounts for the
absence of the stepwise structure in the LS of the 1D LJ gas is also
at work in the present model.

\begin{figure}
\includegraphics[%
  scale=0.32]{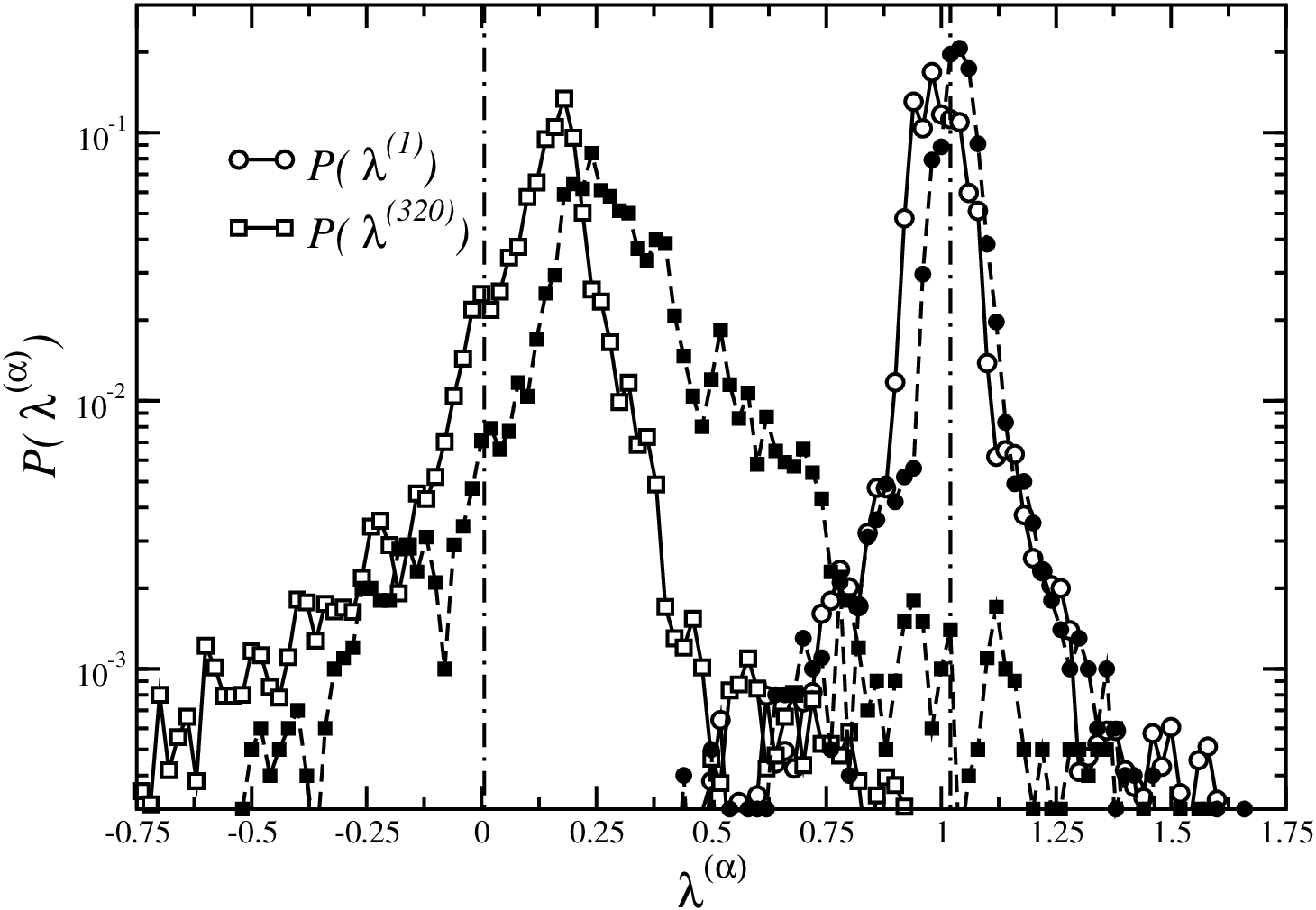}
\caption{\label{cap:Probability-density-function}Probability density function of the instantaneous LEs $\lambda^{\left(1\right)}=0.51$ (circles) and $\lambda^{\left(320\right)}=0.003$ (squares) for $\rho=0.01$. Filled symbols correspond to the results obtained with the potential given by Eq. (\ref{eq:SF}). Average taken over $2\times10^{6}$ integration time-steps. Same $N$ and $T$ values as in Fig. \ref{cap:LS-STS-SF}. Vertical dot-dashed lines indicate the values for $\lambda^{\left(1\right)}$ and $\lambda^{\left(320\right)}$.}
\end{figure}

In Fig. \ref{cap:Probability-density-function} we present the
probability density function for the instantaneous LEs $\lambda^{\left(1\right)}$and
$\lambda^{\left(320\right)}$ for $N=108$ and $\rho=0.01$. It
can be observed that the fluctuations around their mean values increase
as the Lyapunov index $\alpha$ increases, i.e. when going from a
high to a low value of the LE. It is to be observed that the fluctuations
in the $\lambda^{\left(320\right)}$ value are so great that the
tails in the probability density function overlap with those corresponding
to $\lambda^{\left(1\right)}$. These large fluctuations are certainly
a reason why no stepwise structure can be found in the low-$\alpha$
region of the LS.
Other dynamical indicators that we will present
later on are also affected by this behavior of the LS at large $\alpha$
values.
In the same figure results for the potential given by Eq. (\ref{eq:SF})
are also presented; these are quite similar to those obtained with the
STS potential. Thus it is confirmed that the aforementioned fluctuations
are not a result of the discontinuity of the STS potential at $r_c$.
Therefore this potential will be used in the rest of this work.
Other dynamical indicators that we will present
later on are also affected by this behavior of the LS at large $\alpha$
values.

\begin{figure}
\includegraphics[%
  scale=0.3]{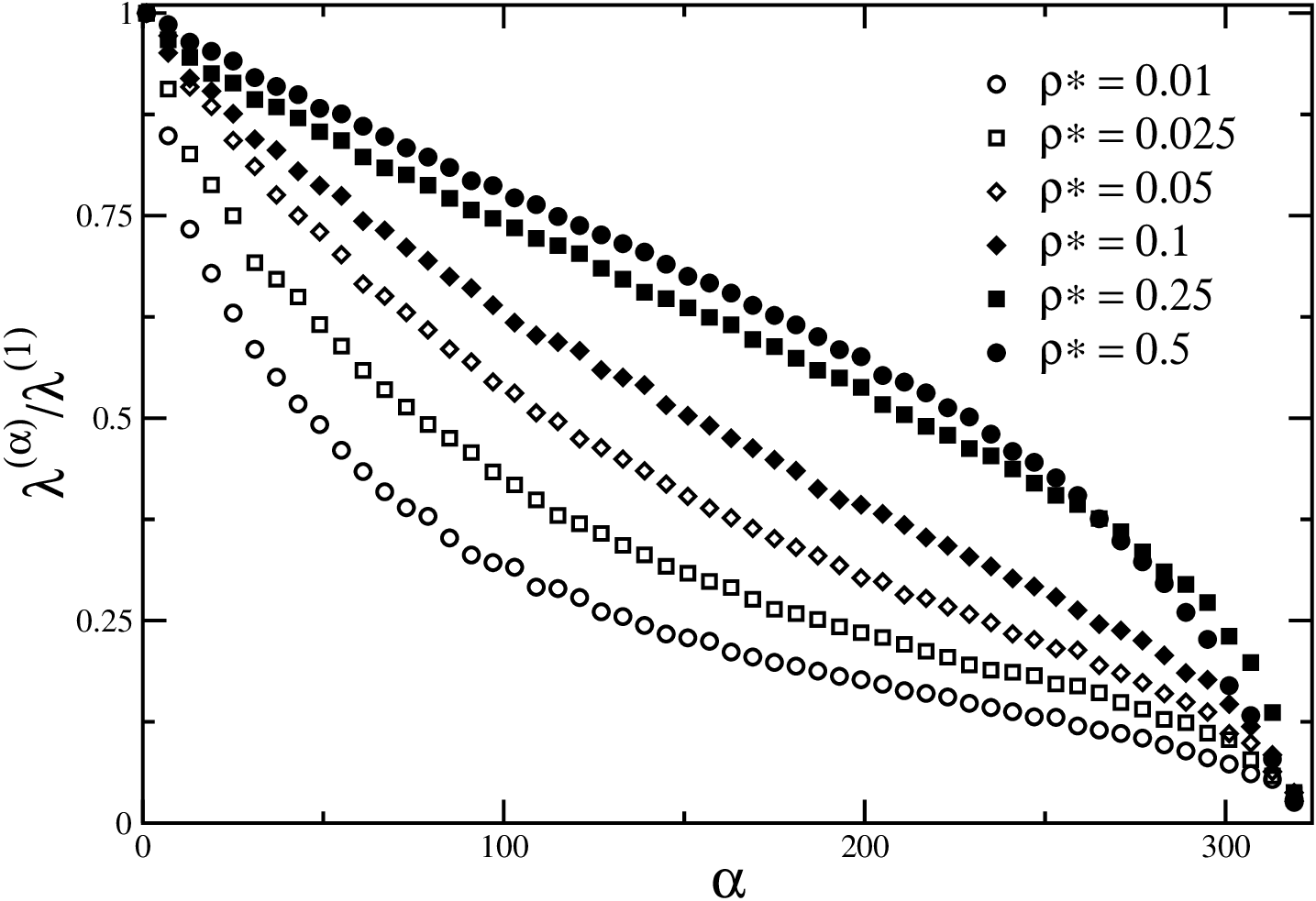}
\caption{\label{cap:Normalized-LS}Normalized LS $\left\{ \lambda^{\left(\alpha\right)}/\lambda^{\left(1\right)}\right\} $ as a function of the Lyapunov index $\alpha$ for all employed values of the reduced density $\rho$ and a reduced temperature of $T=1.5$, with $N=108$.}
\end{figure}

In Fig. \ref{cap:Normalized-LS} we show the normalized LS for $N=108$
and $T=1.5$ computed for the employed range of reduced densities.
 The values of the scalar length of on edge of the cubic simulation cell,
 as well as the values of the LLE corresponding to each value of the
 reduced density, are reported in Table~\ref{tab:table1}. As can be readily appreciated
in Fig. \ref{cap:Normalized-LS}, the LS $\left\{ \lambda^{\left(\alpha\right)}\right\}$
corresponding to the lowest density $\rho=0.01$ can be separated
into two regions. In both of them the LS is a decreasing function
of the Lyapunov index $\alpha$, but in the former the decrease is
more pronounced than in the latter. This bending of the LS has been
observed in a quasi 1D hard-disk gas \cite{Taniguchi}
as well as in the 1D LJ gas \cite{Radons}, and has been related to
the separation of two time scales. To properly explain this point
it is important to remember that each LE indicates a time scale given
by its inverse, so the LS can be considered as a spectrum of time-scales.
The smallest positive LE region of the spectrum is dominated by macroscopic
time and length scale behavior. On the other hand, the opposite
region of the LS is dominated by short time scale behavior, such as
local collision events. As the density increases, the collisions increase
both in frequency and in the number of particles involved; the correlation
among them increases, and becomes increasingly difficult to distinguish
them individually. Thus it is no longer possible to make a separation
of time scales, and so the LS does not present the aforementioned
bending for $\rho\geq0.1$. Although this explanation is plausible
for the LS of our 3D LJ fluid, we also point out that
the bending depicted in Fig. \ref{cap:Normalized-LS} for $\rho=0.01$
(the lowest reduced density value employed) is less pronounced than
the corresponding lowest density instance of the LS of the 1D LJ gas
\cite{Radons}. We interpret this fact as a signature that, even for
the lowest density case, the effect of the dimensionality in the microscopic
events is to enhance the correlation among them, and thus reduce the
separation of time scales that is more evident in the 1D case.
This explanation will be further supported in the next sections.

\begin{table}
\caption{\label{tab:table1} Length $L$ of the cubic simulation cell
 along with the values of the LLE for each reduced density value.}
\begin{tabular}{p{2cm}p{2cm}p{2cm}}
\hline
$\rho$ & $L$ & $\lambda_1$\\
\hline
0.01  & 22.104 & 1.014\\
0.025 & 16.287 & 1.531\\
0.05  & 12.927 & 2.033\\
0.1   & 10.260 & 2.606\\
0.25  & 7.560  & 3.496\\
0.5   & 6.000  & 4.349\\
\hline
\end{tabular}
\end{table}

\section{spatial structure of tangent space perturbations\label{sec:spatial-structure-of-perturbations}}

\subsection{Spatial Lyapunov vector density} \label{sec:SLVD}

\begin{figure}
\includegraphics[%
  scale=0.35]{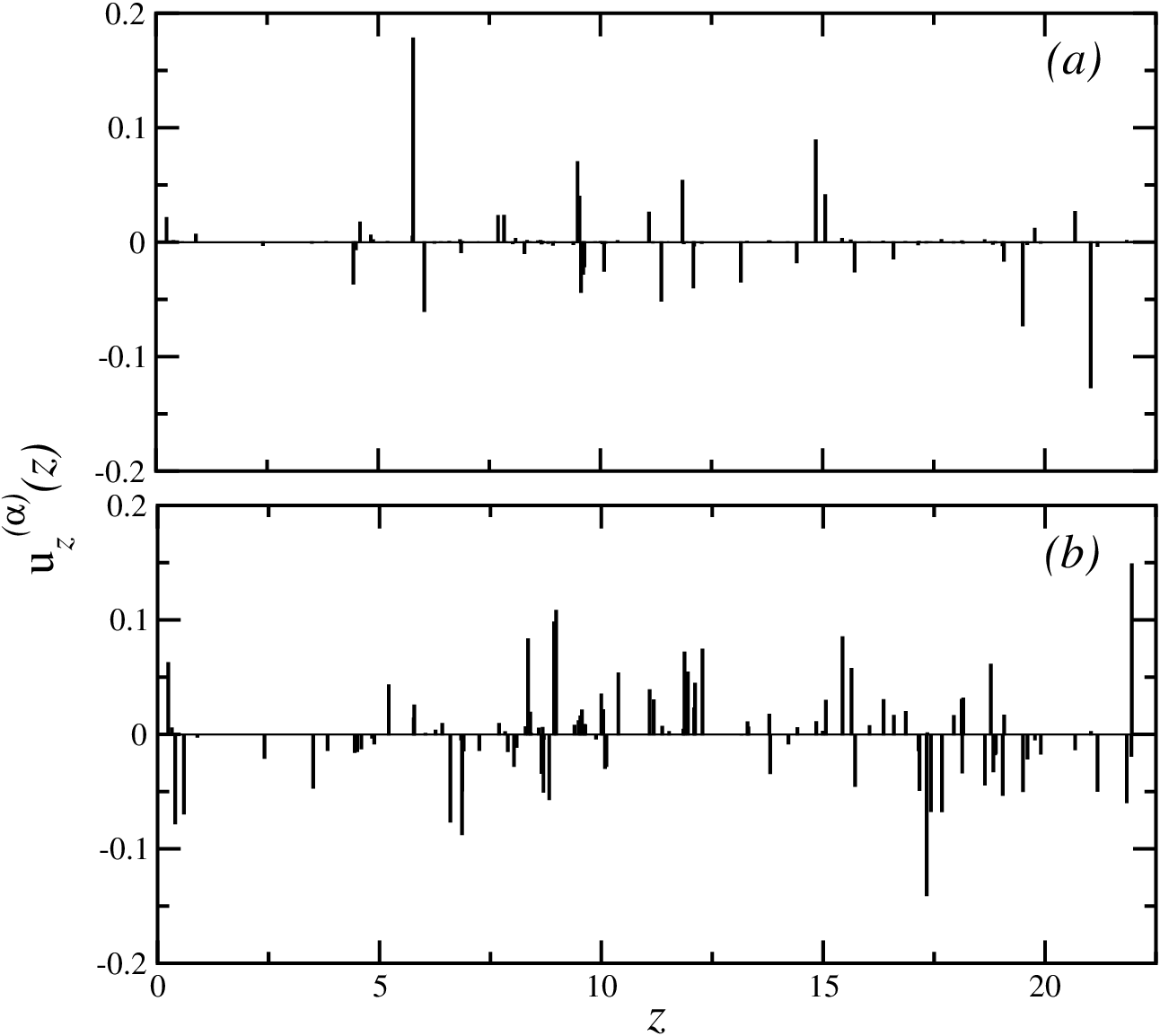}
\caption{\label{cap:Snapshot-LV-density-components}Snapshot of the spatial LV density component $u_{z}^{\left(\alpha\right)}\left(z,t\right)$ along the $z$ axis of the simulation box for a LJ fluid of $N=108$ atoms at a reduced density of $\rho=0.01$ corresponding to the LVs (a) $\alpha=1$ and (b) $\alpha=320$.}
\end{figure}

As discussed in Sec. \ref{sec:Lyapunov-Spectrum}, the strong fluctuations
of the smaller LEs make the perturbations in tangent space extremely
unstable, rendering any coherent structure that may exist in configuration
space difficult to detect. To investigate the possibility that these
structures (LMs) exist in the case of the 3D
STS LJ potential we have to establish a measure of the contribution
of a given LV $\delta\mathbf{\Gamma}^{\left(\alpha\right)}\left(t\right)$
at each point $\mathbf{r}$ of the configuration space, regardless
of which particle makes the contribution to the magnitude of the chosen
LV. We have to remember that the \emph{i}th particle contribution
to the infinitesimal perturbation $\delta\mathbf{\Gamma}^{\left(\alpha\right)}\left(t\right)$
consists of spatial and momentum components. Since the LMs
can be in general considered as Goldstone modes resulting from translational
invariance in coordinate space \cite{deWijn}, we will consider only
the spatial part of the full perturbation component $\delta\mathbf{\Gamma}_{i}^{\left(\alpha\right)}\left(t\right)$.
Thus, in analogy with the definition of microscopic density fluctuations
\cite{Boon}, we define the \emph{spatial LV density} as
\begin{equation}
\mathbf{u}^{\left(\alpha\right)}\left(\mathbf{r},t\right)=\sum_{i=1}^{N}\delta\mathbf{r}_{i}^{\left(\alpha\right)}\delta\left(\mathbf{r}-\mathbf{r}_{i}\right),\label{eq:dens}
\end{equation}
which was first introduced in Ref. \cite{Corr} and is the function to be studied afterwards.
 In Fig. \ref{cap:Snapshot-LV-density-components}
we present a snapshot of a single component $u_{z}^{\left(\alpha\right)}\left(z,t\right)=\sum_{i=1}^{N}\delta z_{i}^{\left(\alpha\right)}\delta\left(\mathbf{r}-\mathbf{r}_{i}\right)$
of the aforementioned spatial density along the $z$ axis of our system.
Since the latter is isotropic, the results are completely analogous
to those corresponding to the $u_{x}^{\left(\alpha\right)}$ and
$u_{y}^{\left(\alpha\right)}$ components projected along their
respective coordinate axes. We can readily appreciate that the spatial
density corresponding to the LLE is more localized than that
corresponding to the LE with $\alpha=320$, which is the lowest index value corresponding to a LV not
related to the space and time translational invariance symmetries of the
system and the associated conserved quantities, the total energy and 
the total momentum \cite{Mareschal}.
A point to be noticed is that the localization of the LLE is
not so clearly defined and strong as in the 1D LJ system \cite{Radons}
or in hard-disk systems \cite{Taniguchi}. This is another indication
that the dimensionality of the system has indeed a strong influence
on the tangent-space dynamics.

Next, we proceed to consider the spatial Fourier transform of $\mathbf{u}^{\,\left(\alpha\right)}\left(\mathbf{r},t\right)$, which can be written as
\begin{eqnarray*}
\widetilde{\mathbf{u}}^{\left(\alpha\right)}\left(\mathbf{k},t\right) & = & \int\mathbf{u}^{\left(\alpha\right)}\left(\mathbf{r},t\right)\exp\left(-\mathrm{i}\mathbf{k}\cdot\mathbf{r}\right)d\mathbf{r}\\
 & = & \sum_{i=1}^{N}\delta\mathbf{r}_{i}^{\left(\alpha\right)}\exp\left[-\mathrm{i}\mathbf{k}\cdot\mathbf{r}_{i}\left(t\right)\right].
\end{eqnarray*}
To proceed further we invoke the \emph{static LV density correlation function} of Ref. \cite{Corr}, defined as
\begin{equation}
\mathcal{C}^{\left(\alpha\right)}\left(\mathbf{k},t\right)\equiv\widetilde{\mathbf{u}}^{\left(\alpha\right)}\left(\mathbf{k},t\right)\widetilde{\mathbf{u}}^{\left(\alpha\right)}\left(-\mathbf{k},t\right),\label{eq:LVDCF}\end{equation}
which in the case of our 3D system is a second rank tensor.
 For our isotropic fluid the Cartesian components 
$C_{\mu\nu}^{\left(\alpha\right)}\left(\mathbf{k},t\right)$
of $\mathcal{C}^{\left(\alpha\right)}\left(\mathbf{k},t\right)$ can be
written in terms of longitudinal $C_{_{L}}^{\left(\alpha\right)}$
and transverse $C_{_{T}}^{\left(\alpha\right)}$ static correlation
functions as
\begin{equation}
C_{\mu\nu}^{\left(\alpha\right)}\left(\mathbf{k},t\right)=\hat{k}_{\mu}\hat{k}_{\nu}C_{_{L}}^{\left(\alpha\right)}\left(\mathbf{k},t\right)+\left(\delta_{\mu\nu}-\hat{k}_{\mu}\hat{k}_{\nu}\right)C_{_{T}}^{\left(\alpha\right)}\left(\mathbf{k},t\right),\label{eq:CompSLVDCF}
\end{equation}
with $\hat{k}_{\mu}=\left(\mathbf{k}/k\right)_{\mu}$. From this last
expression it is immediate to obtain the explicit form of the longitudinal
and transverse static correlation functions as

\begin{eqnarray*}
C_{_{L}}^{\left(\alpha\right)}\left(\mathbf{k},t\right) & = & C_{\mu\nu}^{\left(\alpha\right)}\left(\mathbf{k},t\right)\hat{k}_{\nu}\hat{k}_{\mu}\\
C_{_{T}}^{\left(\alpha\right)}\left(\mathbf{k},t\right) & = & \frac{1}{2}\left(C_{\nu\nu}^{\left(\alpha\right)}\left(\mathbf{k},t\right)-C_{\mu\nu}^{\left(\alpha\right)}\left(\mathbf{k},t\right)\hat{k}_{\nu}\hat{k}_{\mu}\right)
\end{eqnarray*}

To simplify the analysis we will consider a coordinate system such
that the wave vector $\mathbf{k}$ is parallel to the \emph{z} axis.
Then
\begin{equation}
\widetilde{\mathbf{u}}^{\left(\alpha\right)}\left(\mathbf{k},t\right)=\sum_{i=1}^{N}\delta\mathbf{r}_{i}^{\left(\alpha\right)}\exp\left[-\mathrm{i}k_{z}z_{i}\left(t\right)\right].
\end{equation}
For homogeneous systems with translational invariance the Fourier
transformed components of a quantity such as the spatial LV density
$\left\{ \widetilde{u}_{\nu}^{\left(\alpha\right)}\right\} $,
where $\nu=x,y,z$, are uncorrelated \cite{BernePecora}. Thus, our
simplification in no way destroys essential information. The spatial
Fourier spectrum $P_{\nu\nu}^{\left(\alpha\right)}(k_{z},t)\equiv\left|\widetilde{u}_{\nu}^{\left(\alpha\right)}\left(k_{z},t\right)\right|^{2}$
corresponding to each component of the spatial density of $\delta\mathbf{r}_{i}^{\left(\alpha\right)}$
can be readily computed by an algorithm for unequally-spaced data
points \cite{NR} which has been previously applied to the 1D LJ \cite{Radons}
and the 2D WCA systems \cite{WCA}. Furthermore, we observe
that the diagonal components of the static LV density correlation
function correspond to the spatial Fourier spectrum, i.e. $C_{\nu\nu}^{\left(\alpha\right)}\left(k_z,t\right)\equiv P_{\nu\nu}^{\left(\alpha\right)}(k_{z},t)$.
Thus the static longitudinal and transverse correlation functions
can be obtained from the aforementioned power spectra as
\begin{eqnarray*}
C_{_{L}}^{\left(\alpha\right)}\left(k_z,t\right) & = & P_{zz}^{\left(\alpha\right)}(k_{z},t)\\
C_{_{T}}^{\left(\alpha\right)}\left(k_z,t\right) & = & \frac{1}{2}\left(P_{xx}^{\left(\alpha\right)}(k_{z},t)+P_{yy}^{\left(\alpha\right)}(k_{z},t)\right).
\end{eqnarray*}
Finally, since averaging over several spatially equivalent directions will improve the statistics, we take successively the $\mathbf{k}$ vector along the remaining coordinate axes \emph{x} and \emph{y} to obtain two more sets of longitudinal and transverse correlation functions and then average over all sets. The result are longitudinal $C_{_{L}}^{\left(\alpha\right)}(k,t)$ and transverse $C_{_{T}}^{\left(\alpha\right)}(k,t)$ correlation functions independent of the employed coordinate system.

At variance with the hard-core systems in which the patterns resembling
transverse modes do not survive time averaging \cite{Hoover}, the
Fourier spectral techniques so far presented have been quite successful
in the case of the 1D LJ system, and so there is a reasonable possibility
of success in the case of our 3D LJ fluid.

\subsection{Time instability of instantaneous quantities}\label{sec:TIIQ}

\begin{figure}
\includegraphics[%
  scale=0.34]{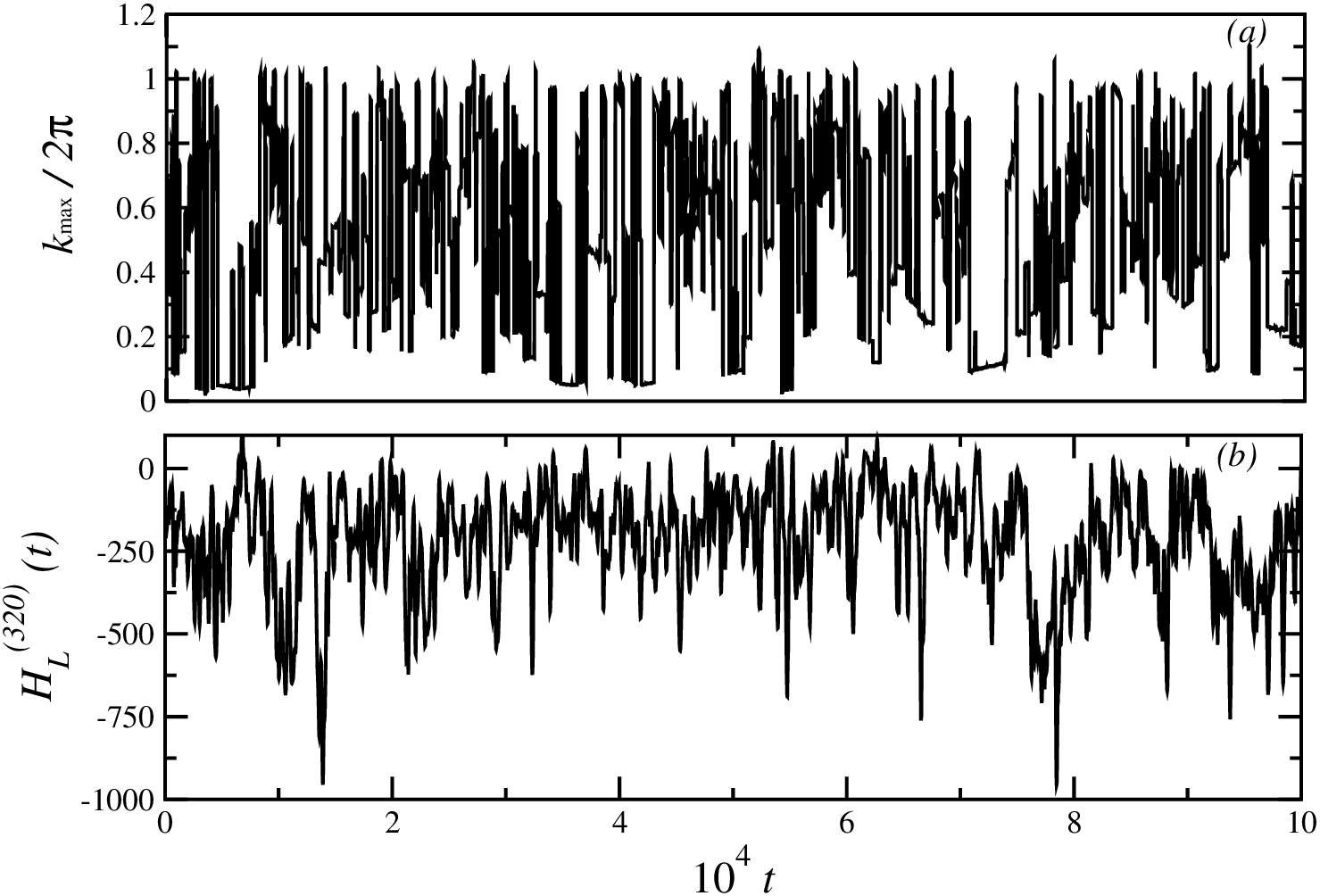}
\caption{\label{cap:Time-evolution-of-peak wavenumber}(a) Time evolution of the peak wave-number $k_{{\mathrm{max}}}$ and (b) spectral entropy $H_{_{L}}^{\left(\alpha\right)}\left(t\right)$ for a Lyapunov index value of $\alpha=320$, for $N=108$, $\rho=0.01$, and $T=1.5$. The behavior of these quantities corresponding to the transverse correlation function $C_{_{T}}^{\left(\alpha\right)}(k,t)$ (not shown) is almost the same as that depicted in the above figure.}
\end{figure}

Due to their mutual interaction, the LMs in all soft potential
systems are only of finite life-time. To investigate their time stability
we present in Fig. \ref{cap:Time-evolution-of-peak wavenumber} the
time evolution of two quantities associated with the longitudinal static
 correlation function $C_{_{L}}^{\left(\alpha\right)}(k,t)$
for the LV $\alpha=320$: the peak wave number $k_{\mathrm{max}}$, which
indicates the position of the highest peak in the spatial Fourier
spectrum $P_{zz}^{\left(\alpha\right)}(k,t)$, and the \emph{spectral
entropy} $H_{_{L}}^{\left(\alpha\right)}(t)$ \cite{SE}, which
measures the $k$ distribution properties of the aforementioned
spectrum. This last quantity was first employed in Ref. \cite{Radons}
in the study of the 1D LJ gas and is defined as
\begin{equation}
H_{_{L}}^{\left(\alpha\right)}\left(t\right)=-\sum_{k}C_{_{L}}^{\left(\alpha\right)}(k,t)\ln C_{_{L}}^{\left(\alpha\right)}(k,t).\label{eq:sentro}
\end{equation}

For the 1D LJ system these quantities show an intermittent behavior,
i.e. large intervals of nearly constant low values (\emph{off state})
are interrupted by short periods of bursts (\emph{on state}) where
they experience large values \cite{Radons}. As can be appreciated
in Fig. \ref{cap:Time-evolution-of-peak wavenumber}(a) for the time
evolution of the peak wave number $k_{\mathrm{max}}$, this behavior is
somewhat different in our case; although there is an alternation between
the on and off states, the mixing among them is so great that there
are no long-lived time intervals for either state, in sharp contrast
with the 1D case. The instant value of the spectral entropy,
presented in Fig. \ref{cap:Time-evolution-of-peak wavenumber}(b),
has a similar behavior as that of $k_{\mathrm{max}}$ in the sense that
it is not easy to identify a correspondence between its temporal
 evolution and that of the on and off states.
A virtually identical behavior is obtained for $k_{\mathrm{max}}$
and $H_{_{T}}^{\left(\alpha\right)}(t)$, but now defined in terms
of $C_{_{T}}^{\left(\alpha\right)}(k,t)$ (not shown),
and for all values of the reduced density. This intermittency in
 the time evolution of the spatial Fourier spectrum of LVs is a
 typical feature of soft-potential systems. However, 
this behavior is greatly increased in comparison to the 1D case.
There are two possible
causes for this difference with the 1D results:
first, the reduced temperature at which we are working
is much higher (an order of magnitude) than that employed in Ref.
\cite{Radons}; second, the dimensionality of our system is also higher,
thus the mixing is easier by this enlargement in the phase space,
as already explained. Now, in order to average out temporal fluctuations,
and thus extract useful information about the collective modes, from
now on we will study the properties of the average spectra
 $\left\langle C_{_{L}}^{\left(\alpha\right)}(k,t)\right\rangle _{t}$
and $\left\langle C_{_{T}}^{\left(\alpha\right)}(k,t)\right\rangle _{t}$,
where $\left\langle\ldots\right\rangle_{t}$ means temporal average.
 Further insight into the temporal dynamics of the LMs can be obtained
 by studying the dynamic LV density correlation function first introduced
 in Ref. \cite{Corr} for the 1D LJ gas. In the Appendix some preliminary results
 of the aforementioned dynamical correlation function applied to our
 3D LJ fluid will be presented.

\subsection{Time-averaged power spectrum} \label{sec:TAPS}

\begin{figure}
\includegraphics[%
  scale=0.45]{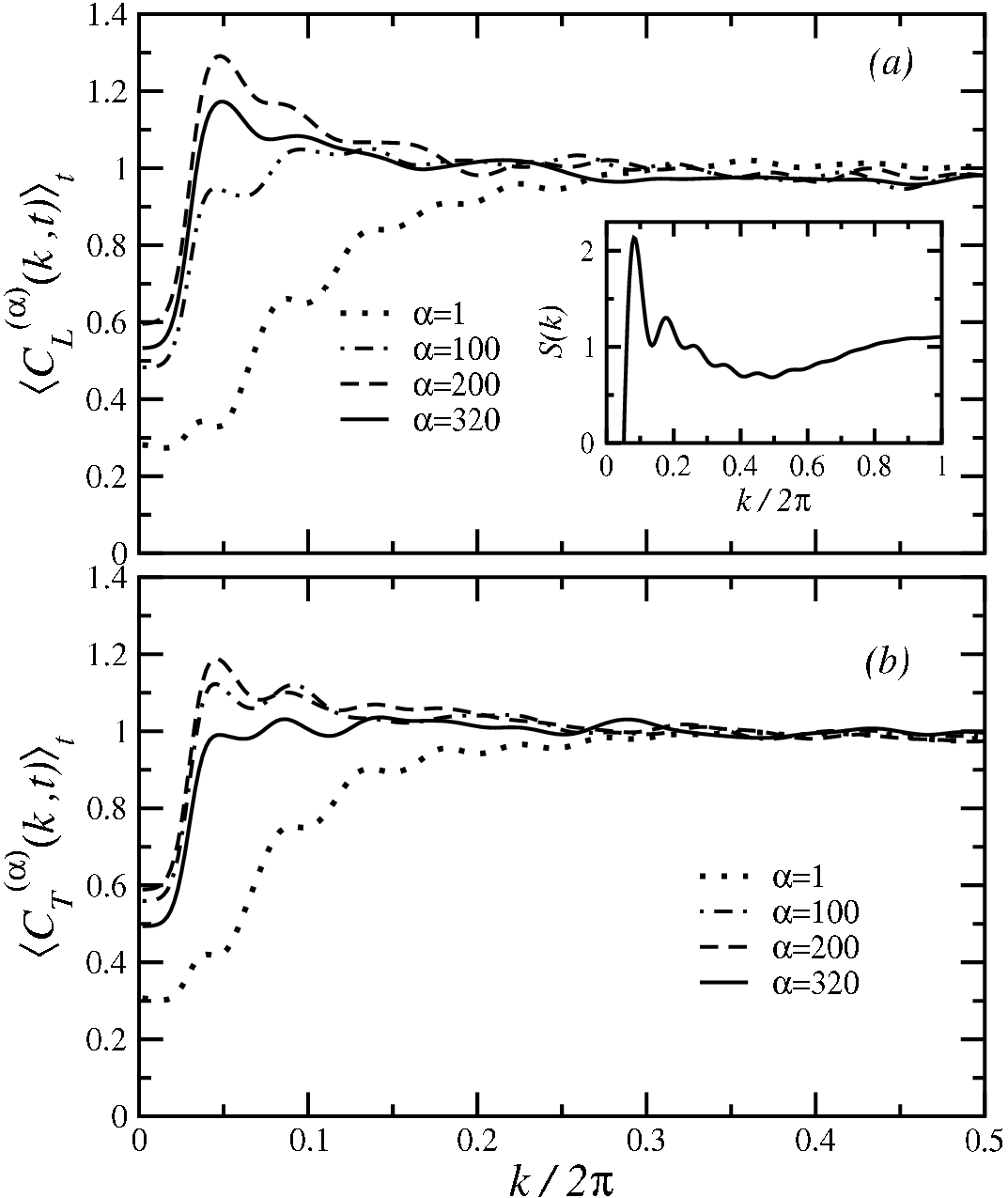}
\caption{\label{cap:Low-density-LT-correlation-functions}(a) Time averaged longitudinal correlation function $\left\langle C_{_{L}}^{\left(\alpha\right)}(k,t)\right\rangle_{t}$ and (b) time averaged transverse correlation function $\left\langle C_{_{T}}^{\left(\alpha\right)}(k,t)\right\rangle_{t}$ for various values of the Lyapunov index $\alpha$, with parameters $N=108$, $T=1.5$, and $\rho=0.01$. In the inset the static structure factor $S(k)$ of the system is plotted.}
\end{figure}

Figures \ref{cap:Low-density-LT-correlation-functions}(a) and 
 \ref{cap:Low-density-LT-correlation-functions}(b) display, for $\rho=0.01$,
$\left\langle C_{_{L}}^{\left(\alpha\right)}(k,t)\right\rangle_{t}$ and 
$\left\langle C_{_{T}}^{\left(\alpha\right)}(k,t)\right\rangle_{t}$
respectively. In both cases we can readily
appreciate that the contribution of the high wave number components
is not small. The instantaneous power spectra are not dominated by
a single peak; rather, several small peaks, which are related to intermediate
length scales, are present, which in turn make significant contributions
to the overall shape of the time averaged correlation functions for
large $k$ values. This feature can be explained by the less pronounced
time-scale separation than in the 1D case, as mentioned
in Sec. \ref{sec:Lyapunov-Spectrum} in relation to the LS. Nevertheless,
the highest value of the time averaged correlation functions is always
dominated by certain low-wave-number components; for the longitudinal
correlation function we observe that the sharp-valued peak in the
spectrum corresponds to diminishing $k_{\mathrm{max}}$ values as the Lyapunov
index $\alpha\rightarrow3N$, i.e. as we go from the region of high
LEs to the region of low LEs. At this point it is important to notice
that this correspondence is not monotonic, since the $k_{\mathrm{max}}$
value of the highest peak is attained at $\alpha=200$, and the height of the corresponding peak slightly
diminishes, although remains well defined, for higher $\alpha$ values. For the transverse correlation
function depicted in Fig. \ref{cap:Low-density-LT-correlation-functions}(b)
the results are similar, except for a very important feature: the
highest value of the spectra is again attained at $\alpha=200$, a
value far away of the region corresponding to the lowest LEs, but
then vanishes as $\alpha\rightarrow3N$. This feature is not consistent
with the existence of well-defined transverse LMs, since
it would be expected that the highest peak in the transverse correlation
function should correspond to small values of $k_{\mathrm{max}}$ in the
$\alpha\approx3N$ regime. In the inset the static structure factor $S(k)$ \cite{Boon}
for the corresponding thermodynamic state is plotted. The highest peak
of this function is located at $k/2\pi\approx0.08$. Now, the position of
the the peak in $\left\langle C_{_{L}}^{\left(\alpha\right)}(k,t)\right\rangle_{t}$
is located at $k_{\mathrm{max}}\approx0.04$. This value suggest that there is no 
obvious direct connection between the short range order of the atoms and
the lowest wavevector peak of the longitudinal LV correlation function.

\begin{figure}
\includegraphics[%
  scale=0.45]{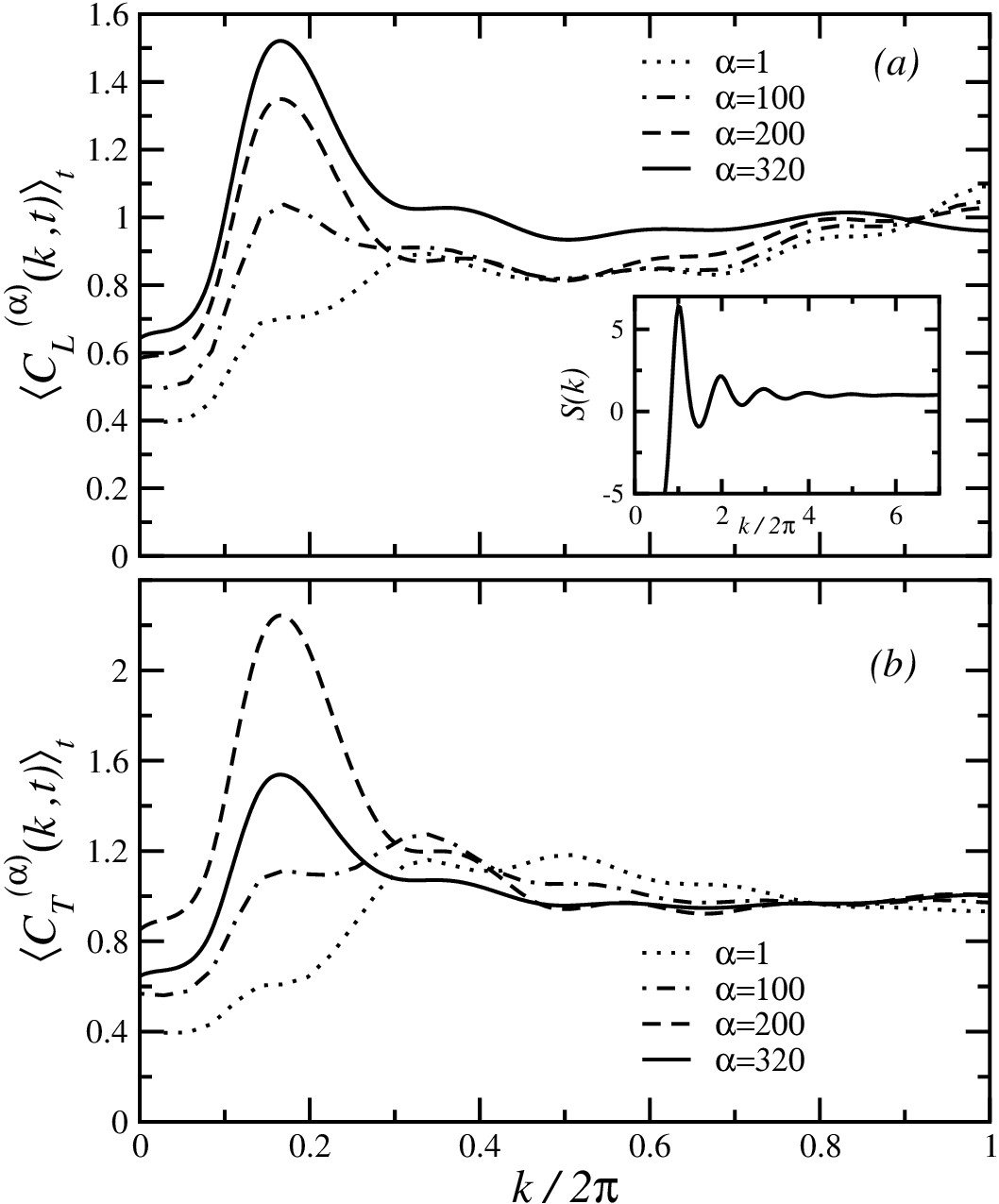}
\caption{\label{cap:High-density-LT-correlation-functions}(a) Time averaged longitudinal and (b) transverse correlation functions for the same parameters as in Fig. (\ref{cap:Low-density-LT-correlation-functions}), but with a reduced density of $\rho=0.5$.}
\end{figure}

The longitudinal and transverse correlation functions corresponding
to the reduced density $\rho=0.5$ are presented in 
Figs. \ref{cap:High-density-LT-correlation-functions}(a) and
\ref{cap:High-density-LT-correlation-functions}(b), respectively. 
A difference with respect to the low density
results is that the highest peak in each power spectra
is broader than in the corresponding
low-density case. This can be interpreted as an indication
that the influence of higher wave numbers is stronger than in the
low-density case. We further observe that the highest peak is located
at a higher wave number value than the corresponding peaks in Figs.
\ref{cap:Low-density-LT-correlation-functions}(a) and (b).
 Now, although at this density there is a higher degree of spatial order
in the atoms of the system, the lowest wavevector peak of $S(k)$, which is plotted in the
inset, has no relation whatsoever with the $k_{\mathrm{max}}$ value 
of the time averaged transverse LV correlation function.
Next we notice that the highest peak of the transverse correlation function
is reached at a value of the Lyapunov index of $\alpha=200$, with
a monotonic decrease for higher $\alpha$ values. This same result
was obtained for the transverse correlation function in the low-density
regime, although in this case the peak remains well defined, in contrast
with the result displayed in Fig. \ref{cap:Low-density-LT-correlation-functions}(b),
 which is almost a flat spectrum. 

\begin{figure}
\includegraphics[%
  scale=0.42]{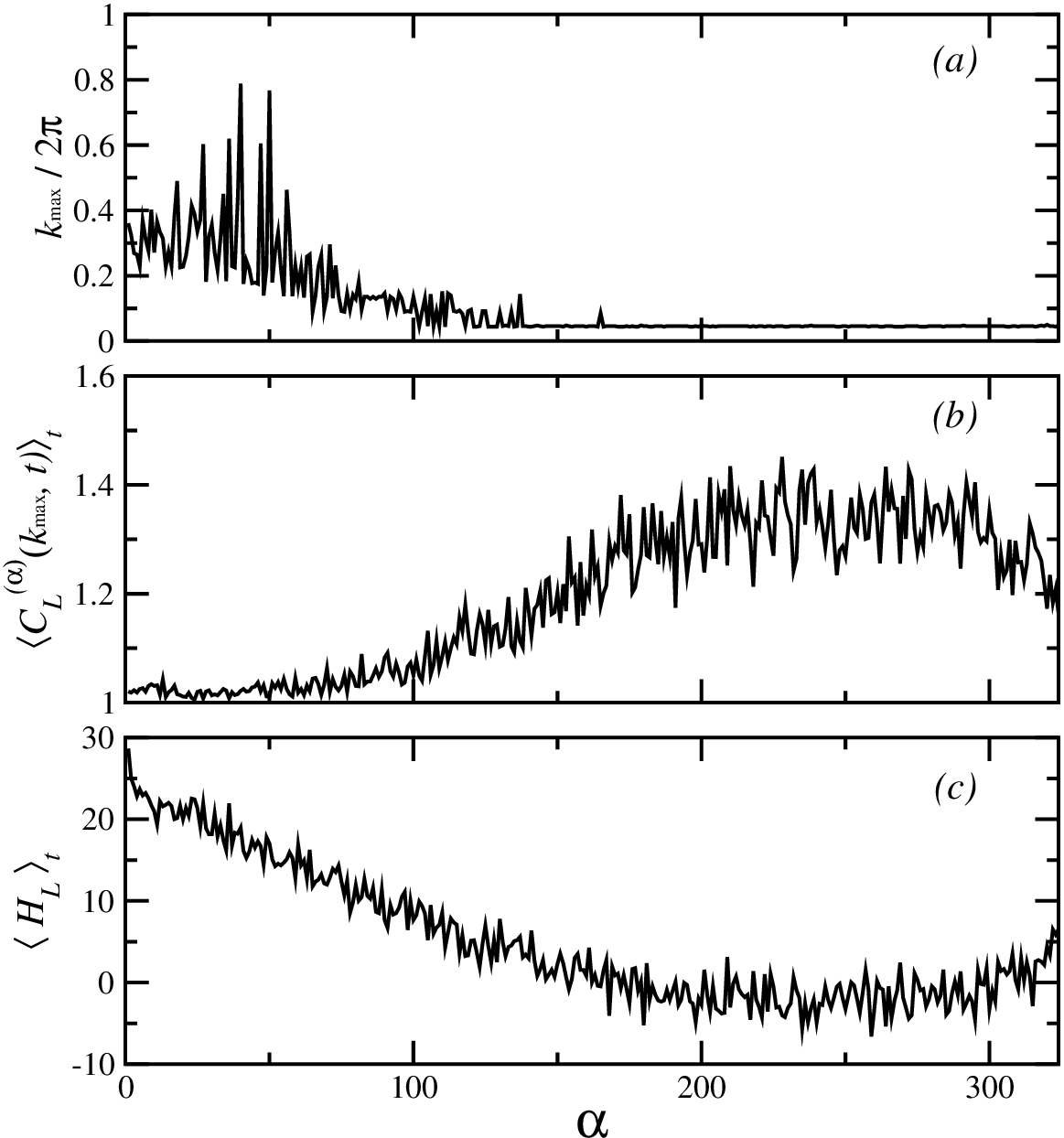}
\caption{\label{cap:Low-density-L-kmaxCmaxH}(a) Wave number $k_{\mathrm{max}}$ of the highest peak in the time-averaged longitudinal correlation function $\left\langle C_{_{L}}^{\left(\alpha\right)}(k,t)\right\rangle_{t}$ for each $\alpha$ value. (b) The height $\left\langle C_{_{L}}^{\left(\alpha\right)}(k_{\mathrm{max}},t)\right\rangle _{t}$ of the highest peak in the time-averaged correlation function. (c) Average spectral entropy $\left\langle H_{_{L}}\right\rangle_{t}$. The parameters are $\rho=0.01$, $N=108$, and $T=1.5$.}
\end{figure}

So far our results are consistent with the existence of longitudinal
LMs. To quantify the properties of these modes in Fig.
\ref{cap:Low-density-L-kmaxCmaxH}(a) we plot $k_{\mathrm{max}}$ versus
the Lyapunov index $\alpha$. It is clear from this figure that, as
$\alpha\rightarrow3N$, $k_{\mathrm{max}}$ diminishes. A feature that
stands out is that, despite the time averaging over a large number
of data points ($10^{6}$ time steps), the obtained values remain
somewhat noisy. Nevertheless it is clear from the figure that the decrease
in the $k_{\mathrm{max}}$ value as $\alpha$ increases is approximately
monotonic, in clear contrast to the 1D case \cite{Radons} in which
a sudden change of $k_{\mathrm{max}}$ from a finite value to zero was
interpreted as a signature of the separation of time scales mentioned
in Sec. \ref{sec:Lyapunov-Spectrum}. Next, in Fig. \ref{cap:Low-density-L-kmaxCmaxH}(b)
the height $\left\langle C_{_{L}}^{\left(\alpha\right)}(k_{\mathrm{max}},t)\right\rangle _{t}$
of the highest peak in the time averaged longitudinal correlation
function is also plotted as function of the Lyapunov index $\alpha$.
On average we can observe a monotonic increase of the peak value as
the Lyapunov index $\alpha$ goes from small to large values, although
a small decrease can be noticed in the $\alpha\approx3N$ region.
Finally, from the definition given in Eq. (\ref{eq:sentro}),
 it is immediate to obtain the average spectral entropy $\left\langle H_{_{L}}\right\rangle _{t}$
which is presented, again as function of $\alpha$, in Fig. \ref{cap:Low-density-L-kmaxCmaxH}(c).
Its value decreases as the Lyapunov index increases. This in turn
means that LVs corresponding to smaller positive LEs are more localized
in Fourier space, i.e. they have more wave-like character than those
corresponding to larger LEs. However, the decrease in the value of
$\left\langle H_{_{L}}\right\rangle _{t}$ as the Lyapunov index increases
is not monotonic. We notice that already for $\alpha=200$ the average spectral
entropy has reached its minimum value, but presents a slight increase as
 $\alpha\rightarrow3N$, which certainly accounts for the decrease in the height of the
 peak in the average spectra of Fig. \ref{cap:Low-density-LT-correlation-functions}(a).
Our tentative conclusion at this point is that the longitudinal LMs
are more vaguely defined than in the 1D case for this density
value.

\begin{figure}
\includegraphics[%
  scale=0.42]{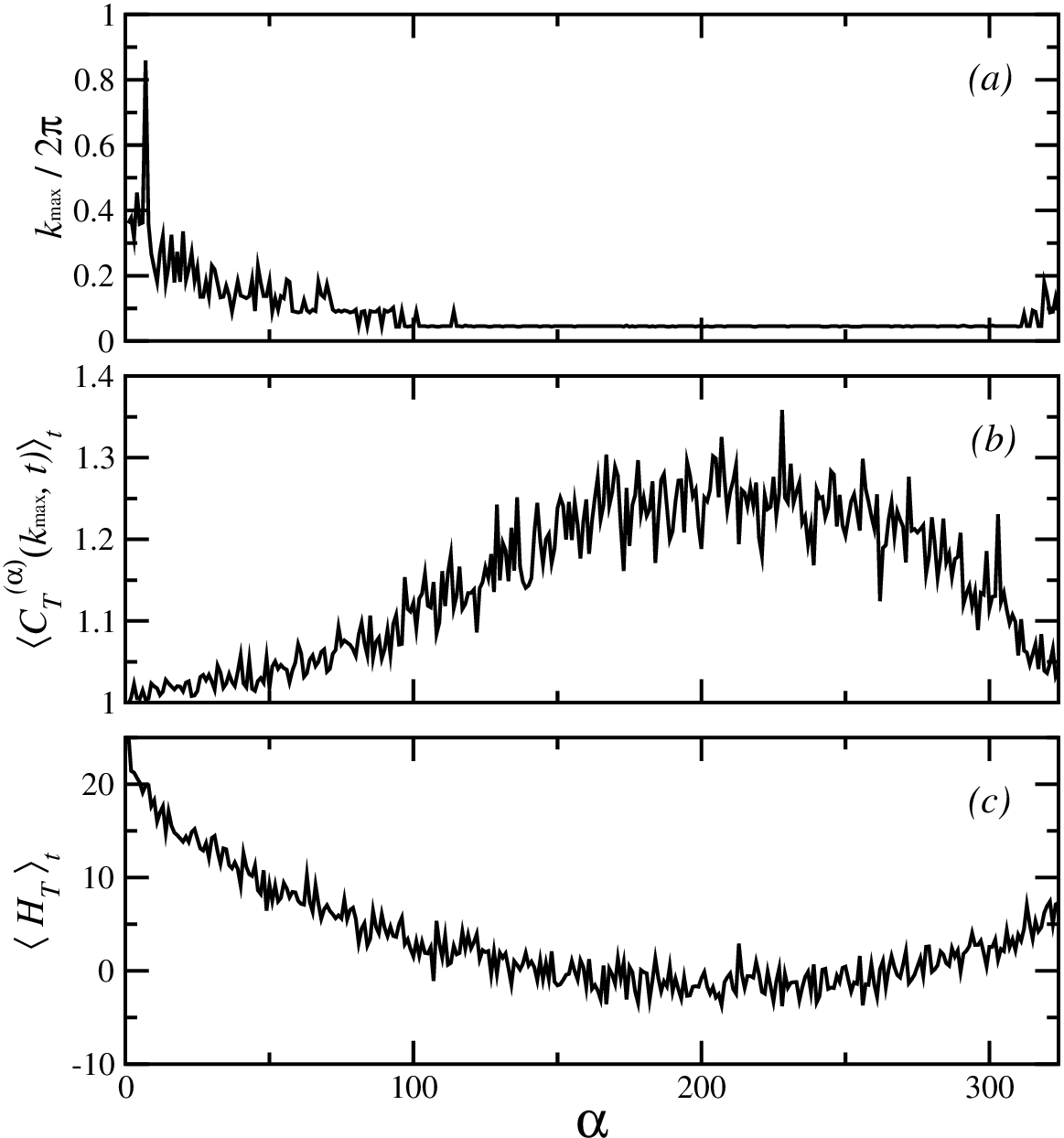}
\caption{\label{cap:Low-density-T-kmaxCmaxH}Same variables as in Fig. (\ref{cap:Low-density-L-kmaxCmaxH}), but corresponding to the time averaged transversal correlation function $\left\langle C_{_{T}}^{\left(\alpha\right)}(k,t)\right\rangle$. Same values of $N$, $\rho$, and $T$ as in Fig. \ref{cap:Low-density-L-kmaxCmaxH}.}
\end{figure}

The results corresponding to the transverse correlation function 
$\left\langle C_{_{T}}^{\left(\alpha\right)}(k,t)\right\rangle_{t}$
are presented in Fig. \ref{cap:Low-density-T-kmaxCmaxH}. We observe
that the $\alpha$ range for which $k_{\mathrm{max}}$ has a small value
is broader compared to the results in Fig \ref{cap:Low-density-L-kmaxCmaxH}(a).
However, $k_{\mathrm{max}}$ has a slight increase in value as $\alpha\approx3N$,
a result which is not consistent with the existence of a transverse
Lyapunov mode for this $\alpha$ range. This conjecture
is supported by the behavior of 
$\left\langle C_{_{T}}^{\left(\alpha\right)}(k_{\mathrm{max}},t)\right\rangle _{t}$
displayed in Fig. \ref{cap:Low-density-T-kmaxCmaxH}(b). This last
quantity has its maximum at $\alpha\approx200$, with a corresponding
minimum of $\left\langle H_{_{T}}\right\rangle _{t}$ at the
same $\alpha$ value, as seen in Fig. \ref{cap:Low-density-T-kmaxCmaxH}(c).
This feature of the spectral entropy is inconsistent with localization
in Fourier space; that is, the wave-like character of the LVs is largely
diminished in the low $\alpha$ region. Taken together these results
make us difficult to unambiguously ascertain the existence
of transverse LMs in the 3D LJ system.

\begin{figure}
\includegraphics[%
  scale=0.42]{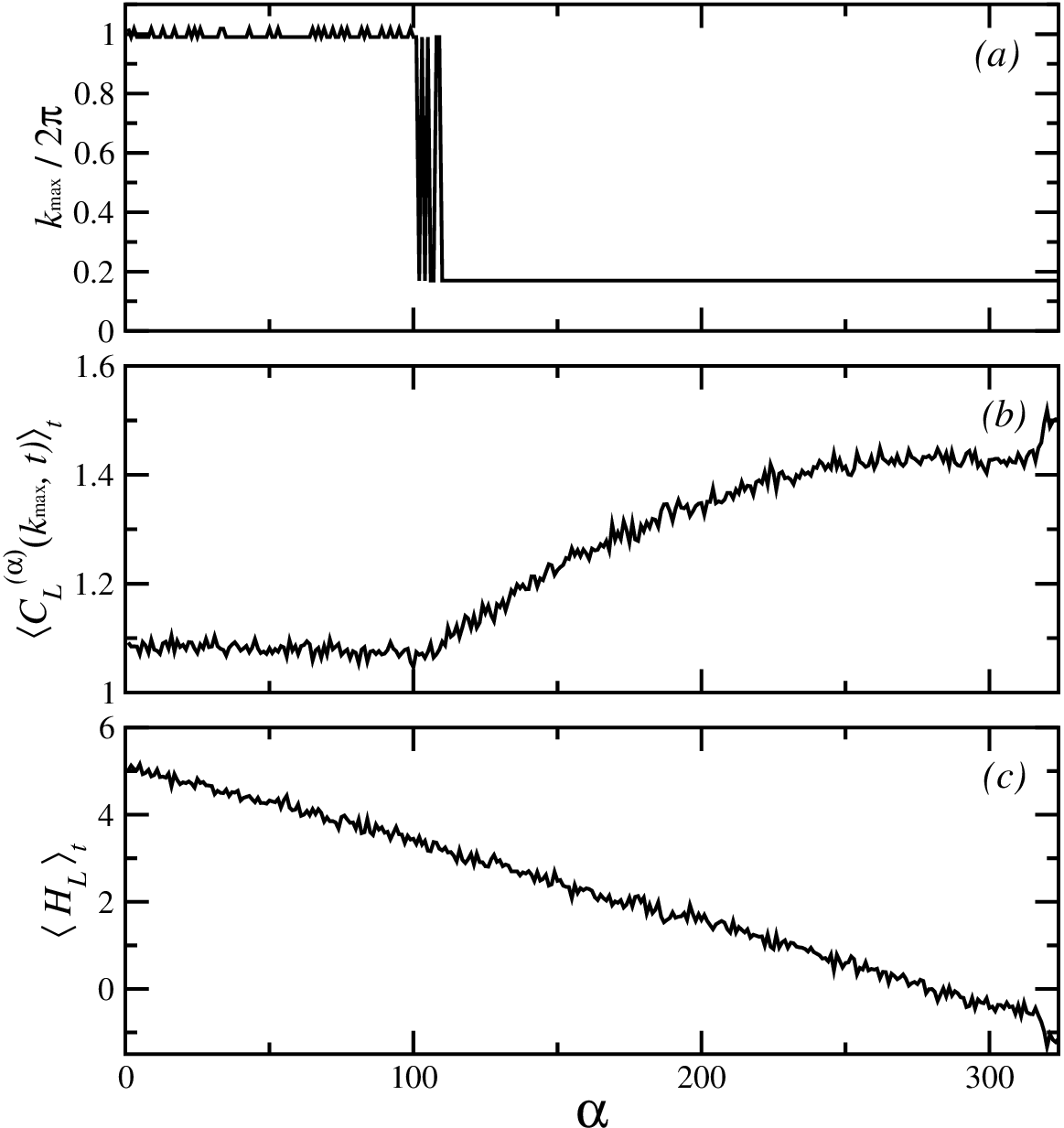}
\caption{\label{cap:High-density-L-kmaxCmaxH}Same variables as in Fig. (\ref{cap:Low-density-L-kmaxCmaxH}) for the time averaged longitudinal correlation function $\left\langle C_{_{L}}^{\left(\alpha\right)}(k,t)\right\rangle$ corresponding to the high-density value $\rho=0.5.$}
\end{figure}

For the case of high reduced density $\rho=0.5$ Fig. \ref{cap:High-density-L-kmaxCmaxH}(a)
presents the results for $k_{\mathrm{max}}$ versus the Lyapunov index
$\alpha$ corresponding to the longitudinal correlation function
 $\left\langle C_{_{L}}^{\left(\alpha\right)}(k,t)\right\rangle_{t}$.
The behavior displayed is very different to that shown in 
Fig. \ref{cap:Low-density-L-kmaxCmaxH}(a).
We first notice that the fluctuations of all quantities are much smaller
than those corresponding to the low-density value. In the present
case the value of $k_{\mathrm{max}}$ stays close to 1 for $\alpha\lesssim100$.
Then, as $\alpha$ further increases and after a small transient interval, drops rather sharply to a value
slightly lower than $k_{\mathrm{max}}=0.2$, which is consistent with that
obtained from Fig. \ref{cap:High-density-LT-correlation-functions}(a).
Next, in Fig. \ref{cap:High-density-L-kmaxCmaxH}(b) we observe that
the value of $\left\langle C_{_{L}}^{\left(\alpha\right)}(k_{\mathrm{max}},t)\right\rangle _{t}$
decreases smoothly from its maximum at $\alpha\approx3N$ down to
its minimum at $\alpha\approx100$. Finally, in Fig. \ref{cap:High-density-L-kmaxCmaxH}(c)
we observe that, at variance with the low-density result, the value
of the average spectral entropy $\left\langle H_{_{L}}\right\rangle_{t}$
decreases monotonically in the whole range of $\alpha$ values, with
a minimum for $\alpha\approx3N$ where the LEs are the smallest possible
ones. From the definition of the spectral entropy we conclude that
the corresponding spectra for these LVs are most significantly dominated
by a few $k$ values. Thus the LMs are more sharply
defined than in the low density case: a very intriguing state of affairs,
 since no relation was found between the spatial order of the atoms as described
by $S(k)$ and the highest peak in the LV power spectrum at any density value.

\begin{figure}
\includegraphics[%
  scale=0.42]{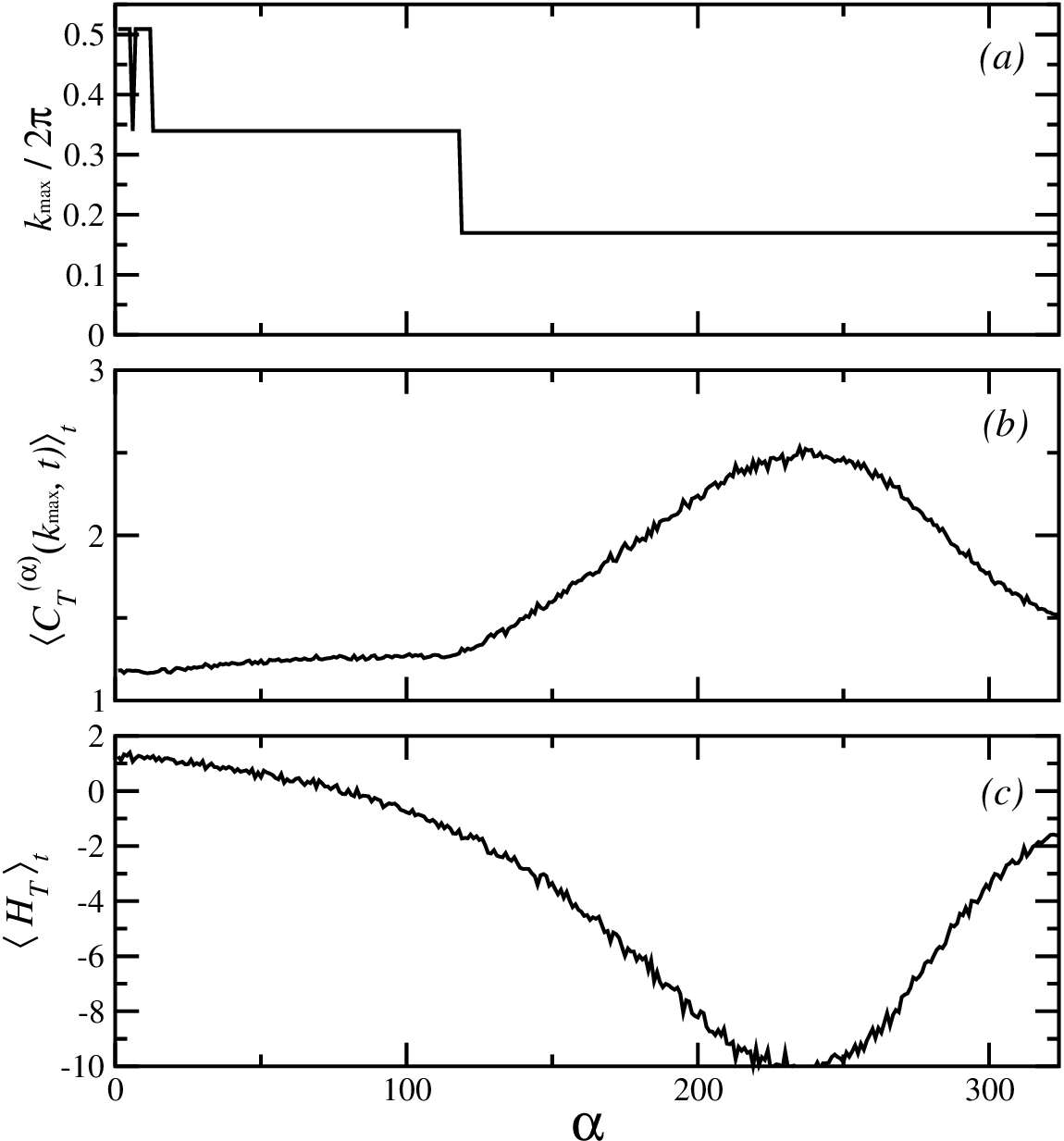}
\caption{\label{cap:High-density-T-kmaxCmaxH}Same variables as in Fig. (\ref{cap:Low-density-L-kmaxCmaxH}) for the time averaged transverse correlation function $\left\langle C_{_{T}}^{\left(\alpha\right)}(k,t)\right\rangle_{t}$ corresponding to the high-density value $\rho=0.5.$}
\end{figure}

The results for the transverse correlation function
 $\left\langle C_{_{T}}^{\left(\alpha\right)}(k,t)\right\rangle _{t}$
for the highest density studied are presented in Fig. \ref{cap:High-density-T-kmaxCmaxH},
 which display a much reduced fluctuation level, but have nevertheless a very similar behavior,
 compared to those in Fig. \ref{cap:Low-density-T-kmaxCmaxH} corresponding to the
 low-density case. That is, the maximum value of 
$\left\langle C_{_{T}}^{\left(\alpha\right)}(k_{\mathrm{max}},t)\right\rangle_{t}$,
which is coincident with the minimum value of $\left\langle H_{_{T}}\right\rangle_{t}$,
is present in a range of $\alpha$ values that is far from the region
in which $\alpha\approx3N$, a result not entirely consistent with the existence
of transverse LMs.

\section{discussion\label{sec:discussion}}

The overall picture that emerges from our results so far indicates
a tangent space dynamics much more complicated than that of hard-core
systems or the 1D LJ fluid. The strong fluctuations in the LEs mentioned
in Sec. \ref{sec:Lyapunov-Spectrum} produce a strong mixture among
modes, so it is unreasonable to expect that results of the spectral
analysis could convey information concerning pure modes. Indeed, the
maximum wave number $k_{\mathrm{max}}$ for both the longitudinal and transverse
correlation functions is highly unstable. In the time interval depicted
in Fig. \ref{cap:Time-evolution-of-peak wavenumber} for the longitudinal
case the $k_{\mathrm{max}}$ value very rarely stays close to zero.
Rather, it seems to wander randomly between zero and one. The instantaneous
value of the spectral entropy, Eq. (\ref{eq:sentro}), also seems to
have a seemingly random time evolution. Finally, there seems to be
no correlation between the time evolution of $k_{\mathrm{max}}$ and 
$H_{_{L}}^{\left(\alpha\right)}(t)$, which was discovered
in the case of the 1D LJ system and which greatly
contributed to a clear-cut definition of the so called on and off
states \cite{Radons}. This is a first indication that the detection
of LMs becomes more difficult than in the corresponding
1D case.

It turns out that the time averaged longitudinal and transverse correlation
functions are the relevant variables from which meaningful information
about the LMs can be obtained. This fact can be understood
in terms of the Zwanzig-Mori formalism, in which the Fourier components
of the fluctuation of a conserved density vary slowly for a small
wave number \cite{BernePecora}; from these ``slow'' variables a meaningful
description is then extracted. For the tangent-space dynamics the
time averaged longitudinal and transverse correlation functions can
be considered as the ``slow'' variables. The results presented in
Figs. \ref{cap:Low-density-LT-correlation-functions} and
 \ref{cap:High-density-LT-correlation-functions}
show that this is indeed the case; the spectra are dominated by low
wave number values, i.e. $k_{\mathrm{max}}\approx0.04$ for $\rho=0.01$
and $k_{\mathrm{max}}\approx0.16$ for $\rho=0.5$, both for longitudinal
and transverse correlation functions, with a corresponding broader
peak in the latter case. In both cases $k_{\mathrm{max}}\approx2\pi/L$,
which is the smallest nonzero wave number allowed by the periodic
boundary conditions used (smaller $k$ values are due to the oversampling
inherent to the method, see Ref. \cite{NR}).
Another point to be noted is that, for our
particular system, the longitudinal correlation function reaches its
maximum value at $\alpha\approx200$, and then its height diminishes slightly,
 but with the peak position $k_{\mathrm{max}}$ unchanged, 
as $\alpha\approx3N$. These results can be attributed to a complicate
mixing of pure modes in the low $k$ regime, which produces the
observed degeneracy of the $k$ value with respect to the $\alpha$
index. Thus the obtained LMs lose their hydrodynamic character
and no dispersion relation $k_{\mathrm{max}}$ vs $\lambda^{\left(\alpha\right)}$
as those observed in the 1D LJ fluid \cite{Radons} and hard-core
systems \cite{Forster} could be detected. Our results even stand in
contrast with those of 2D coupled map lattices, for which a dispersion
relation $\lambda\sim k_{\mathrm{max}}$ indeed exist \cite{RadonsHLMCML}, and even
more sharply with those of the 2D LJ fluid, for which a corresponding
dispersion relation has been claimed to hold, although for
this last model the evidence seems so far to be inconclusive \cite{Corr}. 

The most convincing evidence of the existence of longitudinal LMs
for the low density state comes from the results of Fig. \ref{cap:Low-density-L-kmaxCmaxH}.
First we observe that the average spectral entropy 
attains its minimum at $\alpha\approx200$, a result not entirely
 inconsistent with those of random matrix theory \cite{EckmannGat}.
Next, as $\alpha\rightarrow 3N$, the value of this quantity remains
close to this minimum. Taken together with the already obtained position 
 of the highest peak in the power spectrum in the region $\alpha\approx3N$
 depicted in Fig. \ref{cap:Low-density-LT-correlation-functions}(a),
 these results are compelling evidence of the existence of
 longitudinal LMs. The fuzziness of the obtained values of the
 reported quantities also led us to suppose that the mixing between
 modes is strong.
Up to this point we can affirm that longitudinal LMs do indeed
exist, but are more vague than in the 1D LJ gas and with no hydrodynamical
character at all. On the other hand, our results on the transverse 
correlation function displayed in Fig. \ref{cap:Low-density-T-kmaxCmaxH},
despite their fluctuations, show a tendency that does not allow
us to unambiguously classify them as transverse LMs.

The most important result of this paper was presented in Fig. \ref{cap:High-density-L-kmaxCmaxH}
for $\rho=0.5$. Besides a much reduced fluctuation in the studied
variables, a very defined jump in $k_{\mathrm{max}}$ at $\alpha\approx100$
is observed. The importance of this fact is that, for the 1D LJ gas,
a similar jump was observed, but for a low-density value \cite{Radons}.
In that system this behavior was interpreted in terms as of a separation
of time scales signaled by a bending in the LS. On the contrary, we
obtain a sharply defined jump in $k_{\mathrm{max}}$ at a $\rho$ value
in which the corresponding LS shows no bending, as can be observed
in Fig. \ref{cap:Normalized-LS}.
In order to try to understand
this seemingly puzzling result, we have to remember that, from the
suspected importance of hyperbolicity for the appearance of the 
LMs \cite{EckmannGat}, the main results for the 1D LJ fluid were
obtained in a relatively diluted regime \cite{Radons}, which made
this system somewhat similar to hard-core systems previously studied
\cite{Forster}. However, for $\rho=0.5$ our system is
far from complete hyperbolicity, since the combination of attractive
and repulsive interactions induces strong correlations between the
collision events that makes difficult to separate them from each other.
At the same time, the aforementioned density value is not high enough
to make the effective interaction among atoms similar to that of a
lattice of anharmonic oscillators in which hydrodynamic LMs have
been detected \cite{RadonsFPU}. Thus there is no obvious mechanism that could account for
the sharp jump in $k_{\mathrm{max}}$ when there is no bending in the LS.
 Therefore the role of the alleged separation of
 time scales, inferred from the LS bending for the 1D LJ gas, seems to bear no relation
 to the appearance and phenomenological description of the LMs in
 our 3D LJ fluid.

\section{conclusion\label{sec:conclusions}}

In this paper we have performed a study of the tangent space dynamics of the 3D
LJ fluid in order to investigate the possible existence of the LMs
which are a distinctive feature of the hard-core systems. Our
results indicate that longitudinal modes indeed exist for low and high
reduced density values, and no conclusive evidence of transverse modes
for either density studied. The lack of a dispersion relation between
the LEs and the maximum wave number make both types of modes markedly
different from those already encountered in other systems. The longitudinal
LMs turn out to be much better defined for high values
of the reduced density. Previously only in hyperbolic systems could
LMs be detected at high density values \cite{Forster}.
It is highly plausible that by changing the thermodynamic state of the system
the LMs could present a different behavior than the one
reported in the present communication.

\begin{acknowledgments}
One of the authors (M.R.B.) wishes to acknowledge
 Karen Haneman and D. Casta\~neda-Valle for their comments and suggestions.
 Financial support from Consejo Nacional de Ciencia y Tecnolog\'\i a
 (CONACyT) is also acknowledged.
\end{acknowledgments}

\appendix*

\section{Dynamical LV Correlation Functions}

From the Fourier transform of the spatial LV density $\mathbf{u}^{\,\left(\alpha\right)}\left(\mathbf{r},t\right)$, Eq. (\ref{eq:dens}), we can define the \emph{intermediate two-time correlation function} as
\begin{equation}
\mathcal{F}^{\left(\alpha\right)}\left(\mathbf{k},\tau\right)\equiv\langle\widetilde{\mathbf{u}}^{\left(\alpha\right)}\left(\mathbf{k},t+\tau\right)\widetilde{\mathbf{u}}^{\left(\alpha\right)}\left(-\mathbf{k},t\right)\rangle_t,\label{eq:I2TCF}
\end{equation}
which in the 3D case is a second rank tensor with components $F_{\mu\nu}^{\left(\alpha\right)}\left(\mathbf{k},\tau\right)$. Since the results of Sec. \ref{sec:TAPS} point clearly to the existence of longitudinal LMs, we will concentrate on the longitudinal component of this function. By following the same methodology of Sec. \ref{sec:SLVD} we are led to an expression for the longitudinal component of the form $F_{_{L}}^{\left(\alpha\right)}\left(k_z,\tau\right)=\langle \widetilde{u}_{z}^{\left(\alpha\right)}(k_z,t+\tau)\widetilde{u}_{z}^{\left(\alpha\right)}(-k_z,t)\rangle_t$. Taking {\bf k} parallel to the other coordinate directions and averaging we obtain the final form $F_{_{L}}^{\left(\alpha\right)}\left(k,\tau\right)$. For $\tau=0$ we recover the time average of the static LV correlation function, i.e. $\langle C_{_{L}}^{\left(\alpha\right)}\left(k,t\right)\rangle_t=F_{_{L}}^{\left(\alpha\right)}\left(k,\tau=0\right)$. The dynamical LV correlation function encodes structural as well as temporal correlations, and thus provides more detailed information of the system.
   
\begin{figure}
\includegraphics[%
  scale=0.32]{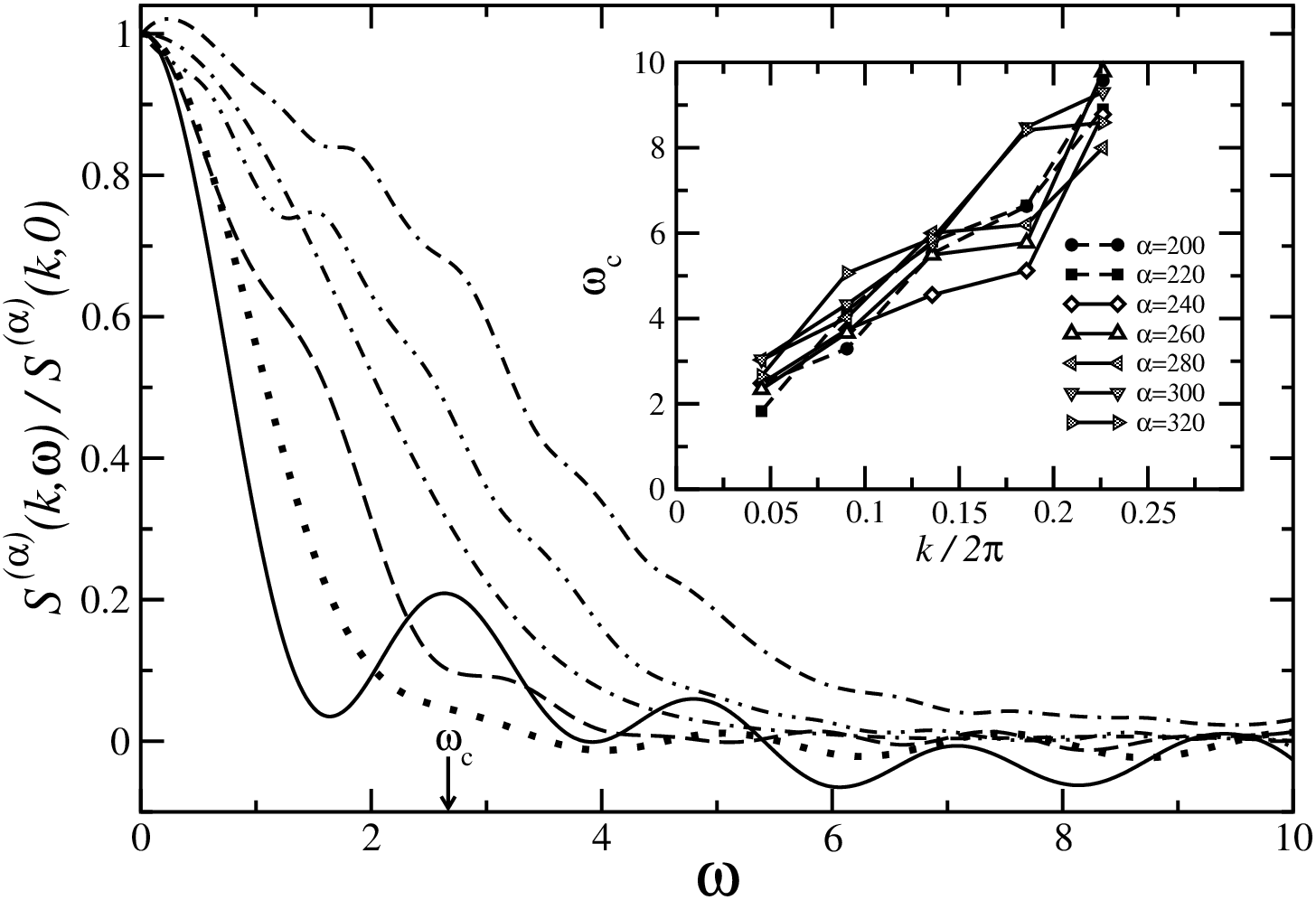}
\caption{\label{cap:Low-density-DLVDCF} Normalized DLVDCF corresponding to the LV $\alpha=320$, with $\rho=0.01$, $T=1.5$, and $N=108$. The curves present results for various values of $k=2\pi n/L$, $n=1,\ldots,6$ from bottom to top. The arrow indicates the position $\omega_c$ of the first peak $n=1$. The inset displays the dispersion relation $\omega^{(\alpha)}(k)$ for different LVs.}
\end{figure}

By Fourier transformation with respect to time we obtain the \emph{dynamical LV density correlation function} (DLVDCF) as $S^{\left(\alpha\right)}\left(k,\omega\right)=(2\pi)^{-1}\int F_{_{L}}^{\left(\alpha\right)}\left(k,\tau\right)\exp(\mathrm{i}\omega\tau)d\tau$. In Fig. \ref{cap:Low-density-DLVDCF} we present the result for $\rho=0.01$ and $\alpha=320$ for the longitudinal DLVCF at various integer multiples of the wavevector $k=2\pi/L$. It can be observed that, besides the central peak, smaller peaks are present as the frequency $\omega$ increases. This result implies that the tangent space dynamics is described by a set of characteristic frequencies, which may have their origin in the lack of timescale separation already advanced in Sec. \ref{sec:Lyapunov-Spectrum}. However, restricting our attention to the lowest frequencies, it is clear that the position $\omega_c$ of the first peak can be unambiguously defined for the lowest $k$ value; a similar identification can be made for other low $k$ values. Thus a dispersion relation $\omega^{\left(\alpha\right)}(k)$ for different $\alpha$ values can be extracted. The result is presented in the inset of the same figure. It is observed that an approximately linear dispersion relation is obtained and thus, besides the characteristic wavevector $k^{\left(\alpha\right)}$ for each LM already determined in Sec. \ref{sec:TAPS}, each mode can also be characterized by a frequency $\omega_c(k^{\left(\alpha\right)})$. The non-vanishing value of $d\omega/dk$ seems to imply propagating wave-like excitations. 

\begin{figure}
\includegraphics[%
  scale=0.32]{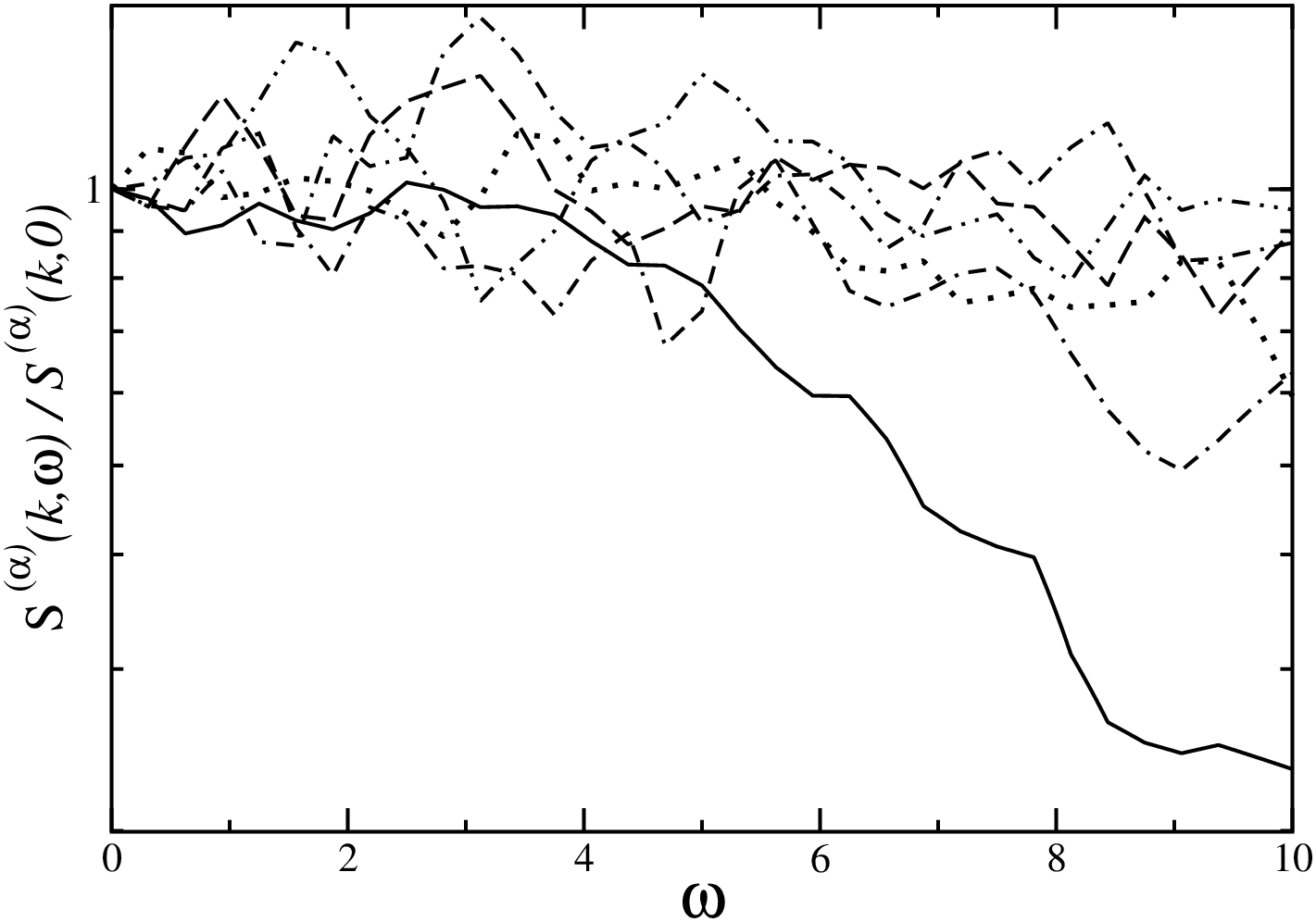}
\caption{\label{cap:High-density-DLVDCF} Normalized DLVDCF corresponding to $\alpha=320$, but for $\rho=0.5$. Same $T$ and $N$ values as in Fig. \ref{cap:Low-density-DLVDCF}.}
\end{figure}

The result for the longitudinal DLVDCF at $\rho=0.5$ is presented in Fig. \ref{cap:High-density-DLVDCF}. In sharp contrast with the low density state no structure can be observed whatsoever at any wavevector number. No characteristic frequency can be easily identified, which implies that no propagation of any wave-like structures occurs at any wavevector number. Thus the tangent space structure described in Fig. \ref{cap:High-density-L-kmaxCmaxH} remains unaltered within the time scales studied. More details of the temporal dynamics in tangent space will be presented in a forthcoming study.


\begin{thebibliography}{10}
\bibitem{KT} Butera P and Caravati G 1987 Phys. Rev. A {\bf36} 962
\bibitem{grav1storder} Torcini A and Antoni M 1999 Phys. Rev. E {\bf59} 2746
\bibitem{phtr2ndorder} Firpo M -C 1998 Phys. Rev. E {\bf57} 6599
\bibitem{Brownian} Romero-Bastida M and Braun E 2002 Phys. Rev. E {\bf65} 036228; Romero-Bastida M 2004 \emph{ibid}. {\bf69} 056204
\bibitem{DC1} Evans D~J, Cohen E~G~D, and Morriss G~P 1990 Phys. Rev. A {\bf42} 5990
\bibitem{DC2} Barnett D~M, Tajima~T, Nishihara~K, Ueshima~Y and Furukawa~H 1996 Phys. Rev. Lett. {\bf76} 1812; Ueshima~Y, Nishihara~K, Barnett~D~M, Tajima~T and Furukawa~H 1997 Phys. Rev. E {\bf55} 3439 (1997); Barnett~D~M and Tajima T 1996 \emph{ibid}. {\bf54} 6084
\bibitem{CommBarnett} Torcini A, Dellago Ch and Posch~H~A 1999 Phys. Rev. Lett. {\bf83} 2676
\bibitem{ReplyBarnett} Barnett~D~M, Tajima~T and Ueshima~Y 1999 Phys. Rev. Lett. {\bf83} 2677
\bibitem{Method1} Evans~D~J and Morriss~G~P 1990 \emph{Statistical Mechanics of Non-equilibrium Liquids} (Academic: New York)
\bibitem{Method2} Gaspard~P 1999 \emph{Chaos, Scattering, and Statistical Mechanics}  (Cambridge: Cambridge University Press)
\bibitem{JRD} Dorfman J~R 1999 \emph{An Introduction to Chaos in Nonequilibrium Statistical Mechanics} (Cambridge: Cambridge University Press)
\bibitem{Milanovic} Milanovi\'{c}~Jj, Posch H~A and Hoover~Wm~H 1998 Mol. Phys. {\bf95} 281
\bibitem{OriginalLM} Posch~H~A and Hirschl R 2000 \emph{Hard Ball Systems and the Lorentz Gas} ed D.~Sz\`asz (Springer: Berlin) p. 279.
\bibitem{EckmannGat} Eckmann J~-P and Gat~O 2000 J. Stat. Phys. {\bf98} 775
\bibitem{TaMoRM} Taniguchi~T and Morriss~G~P 2002 Phys. Rev. E {\bf65} 056202
\bibitem{PeriodicOrbit} Taniguchi~T, Dettmann~C~P and Morriss~G~P 2002 J. Stat. Phys. {\bf109} 747
\bibitem{Mareschal} McNamara~S and Mareschal~M 2001 Phys. Rev. E {\bf64} 051103; Mareschal~M and McNamara~S 2004 Physica D {\bf187} 311
\bibitem{deWijn} de Wijn~A~S and van Beijeren~H 2004 Phys. Rev. E {\bf70} 016207
\bibitem{Hoover} Hoover~Wm~G, Posch~H~A, Forster~Ch, Dellago~C and Zhou~M 2002 J. Stat. Phys. {\bf109} 765
\bibitem{Radons} Yang~H~L and Radons~G 2005 Phys. Rev. E {\bf71} 036211
\bibitem{WCA} Forster~Ch and Posch~H~A 2005 New J. Phys. {\bf7} 32
\bibitem{RadonsPRL} Yang~H~L and Radons~G 2006 Phys. Rev. Lett. {\bf96} 074101
\bibitem{RadonsHLMCML} Yang~H~L and Radons~G 2006 Phys. Rev. E {\bf73} 016202
\bibitem{RadonsDynBehavHLMCML} Yang~H~L and Radons~G 2006 Phys. Rev. E {\bf73} 016208
\bibitem{RadonsFPU} Yang~H~L and Radons~G 2006 Phys. Rev. E {\bf73} 066201
\bibitem{AT} Allen~M~P and Tildesley~D~J 1987 \emph{Computer Simulations of Liquids} (Oxford: Oxford University Press)
\bibitem{SFord} Stoddard~S~D and Ford~J 1973 Phys. Rev. A {\bf8} 1504
\bibitem{BSmit} Smit~B 1992 J. Chem. Phys. {\bf96} 8639
\bibitem{Oseledec} Oseledec~V~I 1968 Trans. Mosc. Math. Soc. {\bf19} 197
\bibitem{Benettin} Benettin~G, Galgani~L and Strelcyn~J~M 1976 Phys. Rev. A {\bf14} 2338
\bibitem{Shimada} Shimada~I and Nagashima~T 1979 Prog. Theor. Phys. {\bf61} 1605
\bibitem{Johnson} Johnson~R~A, Palmer~K~J and Sell~G~R 1987 SIAM J. Math. Anal. {\bf18} 1
\bibitem{Orszag} Goldhirsch~I, Sulem~P~L and Orszag~S~A 1987 Physica D {\bf27} 311
\bibitem{Ershov} Ershov~S~V and Potapov~A~B 1998 Physica D {\bf118} 167
\bibitem{Diego} Szendro~I~G, Paz\'o~D, Rodr\'\i guez~M~A and L\'opez~J~M 2007 Phys. Rev. E {\bf 76} 025202
\bibitem{Ginelli} Ginelli~F \emph{et. al.} 2007 Phys. Rev. Lett. {\bf 99} 130601
\bibitem{ER} Eckmann~J~-P and Ruelle~D 1985 Rev. Mod. Phys. {\bf 57} 617
\bibitem{Diatomic} Yang~H~L and Radons~G 2007 Phys. Rev. Lett. {\bf96} 164101
\bibitem{XYmodel} Yang~H~L and Radons~G 2008 Phys. Rev. E {\bf77} 016203
\bibitem{cpr1} Dettmann~C~P and Morriss~G~P 1996 Phys. Rev. E {\bf53} R5545
\bibitem{cpr2} Ruelle~D 1999 J. Stat. Phys. {\bf95} 393
\bibitem{Taniguchi} Taniguchi~T and Morriss~G~P 2003 Phys. Rev. E {\bf68} 046203
\bibitem{Boon} Boon~J~P and Yip~S 1991 \textit{Molecular Hydrodynamics} (New York: Dover)
\bibitem{Corr} Radons~G and Yang~H~L \textit{Preprint} nlin.CD/0404028.
\bibitem{BernePecora} Berne~B~J and Pecora~R 2000 \emph{Dynamic Light Scattering} (New York: Dover)
\bibitem{NR} Press~W~H, Teukolsky~S~A, Vetterling~W~T and Flannery~B~P 1992 \textit{Numerical Recipes in Fortran 77} (Cambridge: Cambridge University Press)
\bibitem{SE} Livi~R, Pettini~M, Ruffo~S, Sparpaglione~M and Vulpiani~A 1985 Phys. Rev. A {\bf31} 1039
\bibitem{Forster} Forster~C, Hirschl~R, Posch~H~A and Hoover~Wm~H 2004 Physica D {\bf187} 294

\end{thebibliography}
\end{document}